\documentclass[useAMS,onecolumn]{mn2e}

\newif\ifAMStwofonts

\usepackage{subfigure}
\usepackage[dvips]{graphicx}
\usepackage{amssymb}	% Extra maths symbols
\usepackage{epsfig}
\usepackage{color}

\def\pmb#1{\mbox{\boldmath$#1$}}

\def\gtsim {\gtrsim}
\def\ltsim {\lesssim}

\def\be{\begin{equation}}
\def\ee{\end{equation}}
\def\be{\begin{eqnarray}}
\def\ee{\end{eqnarray}}

\def\pmbmt#1{\pmb{\sf #1}}
\def\rmi{{\rm i}}

\begin{document}

\title[Thermal Tides in Hot Jupiters]{Thermal Tides in Rotating Hot Jupiters}

\author[U. Lee, D. Murakami]{
Umin Lee\thanks{e-mail:lee@astr.tohoku.ac.jp} and
Daiki Murakami\thanks{e-mail:d.murakami@astr.tohoku.ac.jp}
\\
Astronomical Institute, Tohoku University, Sendai, Miyagi 980-8578, Japan
}

\date{Accepted XXX. Received YYY; in original form ZZZ}
\pubyear{2019}

\maketitle

\begin{abstract}
We calculate tidal torque due to semi-diurnal thermal tides in rotating hot Jupiters,
taking account of the effects of radiative cooling in the envelope and of the planets rotation
on the tidal responses.
We use a simple Jovian model composed of a nearly isentropic convective core and a thin radiative envelope.
To represent the tidal responses of rotating planets, we employ series expansions
in terms of spherical harmonic functions $Y_l^m$ with different $l$s for
a given $m$.
For low forcing frequency, there occurs frequency resonance between the forcing and 
the $g$- and $r$-modes in the envelope and inertial modes in the core.
We find that the resonance enhances the tidal torque, and that
the resonance with the $g$- and $r$-modes produces broad peaks and that with the inertial modes very sharp peaks, 
depending on the magnitude of the non-adiabatic effects associated with
the oscillation modes.
We also find that the behavior of the tidal torque as a function of the forcing frequency
(or period) is different between prograde and retrograde forcing, particularly for long forcing periods
because the $r$-modes, which have long periods, exist only on the retrograde side.
\end{abstract}


\section{Introduction}

It is well known that equilibrium and dynamical gravitational tides play various important roles in the dynamics of
planetary systems and in binary systems of stars 
(e.g., Goldreich \& Soter 1966; Zahn 1977; Press \& Teukolsky 1977;
Savonije \& Papaloizou 1984, 1997; Lai 1997; Witte \& Savonije 2002; Ivanov \& Papaloizou 2007; 
Ogilvie \& Lin 2004; Ogilvie 2014; Fuller \& Lai 2013).
Because of energy dissipations accompanied by the
tidal responses raised in stars and planets, the gravitational tides can be a cause of 
synchronization between the rotation and the orbital motion, and of circularization of 
the orbits.

Hot Jupiters are giant gas planets orbiting the central star at distances as close as less than about 0.05A.U..
Observationally it is suggested that hot Jupiters have larger radii compared to Jovian planets of the similar mass and age orbiting far from the central stars (e.g., Jermyn, Tout \& Ogilvie 2017).
Baraffe et al (2003), for example, have numerically shown that the planets would inflate to the extent observationally determined if there exists an effective heating source deep in the atmosphere.
As a possible heating mechanism, tidal heating in the planets was suggested by Bodenheimer, Lin, Mardling (2001).
In the case of hot Jupiters closely orbiting the central star, however, the gravitational tides are
strong enough to make the spin and orbital motion synchronized in a timescale much shorter than the ages of the planets, which suggests that another mechanism is needed that keeps the spin and orbital motion
asynchronous.
For such a mechanism for hot Jupiters, Arras \& Socrates (2010)  
suggested thermal tides, which would be operative because of strong irradiation by the central star.
The periodic alternations of day and night sides on the planets may bring about semi-diurnal density perturbations in the planets, which could cancel the effects of the density perturbations produced by the gravitational tides.
If this is the case, the spin and orbital motion would be kept asynchronous 
so that the tidal heating can be a viable mechanism to inflate the planets (e.g., Arras \& Socrates 2010). 

Following the suggestion by Arras \& Socrates (2010),
Auclair-Desrotour \& Leconte (2018) computed the tidal torque caused by
semi-diurnal thermal tides, taking into considerations the effects of energy dissipations caused by radiative cooling in the envelope and those of the planets rotation on the tidal responses.
For hot Jupiters, they used very simple models, composed of a convective core and a thin radiative envelope.
The convective core has the structure corresponding to that of a polytrope of the index $n=1$ and
the envelope is nearly isothermal.
To represent the tidal responses in rotating planets, Auclair-Desrotour \& Leconte (2018)
employed series expansions in terms of the Hough functions, defined in the traditional approximation for the perturbations in rotating bodies (e.g., Lee \& Saio 1997).
They found that the tidal torque due to thermal tides is affected by frequency resonance between
the tidal forcing and the $g$-modes in the envelope, changing its sign as a function of the forcing period.
They also discussed the possibility to produce differential rotation at the sites where the tide exerts
strong local torques.

We revisit the problems of the semi-diurnal thermal tides in rotating hot Jupiters, 
following Auclair-Desrotour \& Leconte (2018).
In this paper, however, we do not employ the traditional approximation to represent tidal responses
in the rotating planets.
Instead, we simply use series expansions
in terms of spherical harmonic functions $Y_l^m(\theta,\phi)$ with different $l$s for a given $m$ (see, e.g., Lee \& Saio 1986, 1987).
We also assume the convective core of the rotating planets is nearly isentropic 
(e.g., Stevenson \& Salpeter 1977a,b; Stevenson 1979) so that the core can support propagation of
inertial modes.
%{\bf 
In this paper, we discuss for rotating Jovian planets tidally excited low frequency modes such as $g$-modes, inertial modes and $r$-modes.
The $g$-modes are internal gravity waves propagating in stably stratified regions expected for the radiative envelope
of the planet.
The latter two are rotationally induced modes for which the Coriolis force is the restoring force.
There exist prograde and retrograde inertial modes but $r$-modes exist only on the retrograde side 
(Papaloizou \& Pringle 1978; see also Greenspan 1969; Unno et al 1989). 
$r$-modes result from the interaction between the radial component of the 
vorticity and the Coriolis force and form a subclass of inertial modes.
Note that although inertial modes propagate in nearly isentropic regions,
the $r$-modes discussed in this paper are those propagating in the radiative envelope.
See also the 5th paragraph in \S 3.
%}
The method of solution used in this paper is described in \S 2 and the numerical results are given in \S 3.
We conclude in \S 4.

\section{Basic Equations}

\subsection{Equilibrium Model}

Following Arras \& Socrates (2010) and Auclair-Desrotour \& Leconte (2018), to compute tidal responses of 
strongly irradiated Jovian planets, we use a
simple model, composed of an irradiated thin isothermal atmosphere and a nearly isentropic
convective core.
Such Jovian models are obtained by integrating the equations for hydrostatic equilibrium %{\bf 
(e.g., Clayton 1968)%}
\be
{dp\over dr}=-\rho{GM_r\over r^2},
\ee
\be
{dM_r\over dr}=4\pi r^2\rho,
\ee
with the analytic equation of state given by
\be
\rho(p)=e^{-p/p_b}{p\over a^2}+\left(1-e^{-p/p_b}\right)\sqrt{p\over K_c},
\ee
where $p$, $\rho$, $M_r$, and $G$ are respectively the gas pressure, the mass density, the mass within the sphere of radius $r$ and the gravitational constant, and $p_b$ is the pressure at the base of the stably stratified layer (radiative atmosphere)
and
\be
K_c=GR_J^2, \quad a^2=\sqrt{p_bK_c},
\ee
where $a$ is the isothermal sound velocity and $R_J$ is the radius of Jupiter.
In the convective core where $p\gg p_b$, we obtain
\be
p\approx K_c\rho^{\Gamma} \quad {\rm with} \quad \Gamma=2,
\ee
and hence the square of the Brunt-V\"ais\"al\"a frequency 
\be
N^2=-g\left({d\ln\rho\over dr}-{1\over\Gamma_1}{d\ln p\over dr}\right)\equiv -gA\approx 0 \quad {\rm for}\quad \Gamma_1\equiv \left({\partial\ln p\over\partial\ln\rho}\right)_{ad}=2,
\ee
where $g=-GM_r/r^2$, and $A$ is known as the Schwarzschild discriminant.
On the other hand, in the radiative layers where $p\ll p_b$, we have
\be
p\approx a^2\rho,
\ee
and
\be
N^2\approx 
(\Gamma_1-1){g^2\over c^2}, \quad c^2=\Gamma_1{p\over\rho},
\ee
and $c$ is the adiabatic sound speed.
Since $p/\rho\approx a^2 \propto T$ for an ideal gas, 
the temperature $T$ is constant for a constant $a$, suggesting an isothermal atmosphere.

In this paper, we use $p_b=100{\rm bar}=10^8{\rm dyn/cm^2}$ and set the outer boundary $R_e$ at
$p=0.01{\rm dyn/cm^2}$ and define the planet's radius $R=R_e/1.01$.
We use $M=0.7M_J$ with $M_J$ being the mass of Jupiter and the radius $R=9.31\times10^9$cm.
Figure 1 is the propagation diagram of the Jovian model we use, where $N^2$
and $L_l^2\equiv l(l+1)c^2/r^2$, normalized by $GM/R^3$, are plotted for $l=2$ versus $\log_{10} p$.
%{\bf 
Note that $L_l$ is called Lamb frequency and indicates the local lower limit of frequency for the
propagation of sound waves.%}
We have $N^2\approx 0$ for $\Gamma_1=2$ in the convective core, and $N^2>0$ in the radiative envelope.
Since $g$-modes of the frequency $\omega$ propagate in the regions where $\omega< N$ and $\omega < L_l$,
Figure 1 indicates that
the $g$-modes that propagate in the radiative envelope of the model have frequencies $\omega\ltsim 0.1\sqrt{GM/R^3}$.

\begin{figure}
\resizebox{0.45\columnwidth}{!}{
\includegraphics{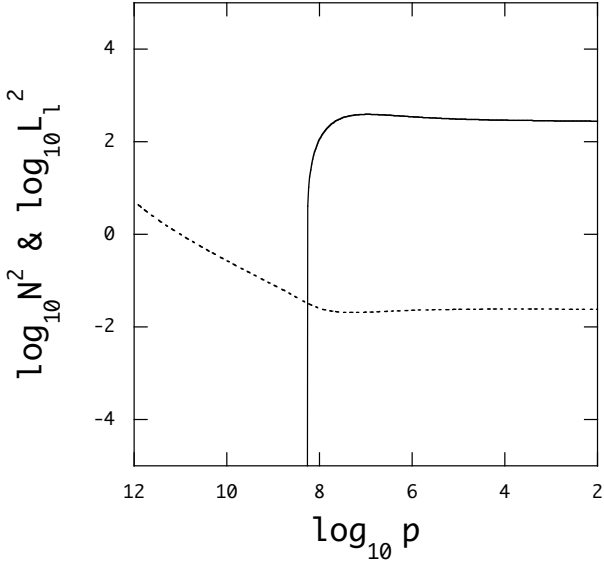}}
\caption{Square of the Brunt-V\"ais\"al\"a frequency $N$ (solid line) and the Lamb frequency $L_l$ for $l=2$ (dotted line) plotted as a function of $\log_{10}p$ for the Jovian model, where $N^2$ and $L_l^2$ are normalized by ${GM/R^3}$ with $M$ and $R$ being the mass and radius of the model.
}
\end{figure}

\subsection{Perturbed Basic Equations}

We treat tidal responses as small amplitude perturbations of the planet.
We assume that the planet is uniformly rotating at the angular speed $\Omega$ and is orbiting
the host star in a circular orbit with the frequency $\Omega_{\rm orb}=\sqrt{(M_*+M)/a_*^3}$,
where $a_*$ denotes the semi-major axis, and $M_*$ and $M$ are the mass of the host star and the planet, respectively.
Assuming that the perturbing tidal potential $\Phi_T$ due to the
host star depends on the time as $\Phi_T\propto e^{\rmi\omega t}$ with $\omega$ being the forcing frequency, 
the basic equations for the small amplitude tidal responses of the planet may be governed by 
%{\bf 
(see, e.g., Unno et al 1989)%}
\be
-\rho\omega^2\pmb{\xi}+2\rmi\rho\omega\pmb{\Omega}\times\pmb{\xi}-{\rho'\over\rho}\nabla p+\nabla p'=-\rho\nabla\Phi_T,
\label{eq:eqmot}
\ee
\be
\rho'+\nabla\cdot\left(\rho\pmb{\xi}\right)=0,
\label{eq:cont}
\ee
\be
\rmi\omega\rho T\delta s=\left(\rho\epsilon\right)'-\nabla\cdot\pmb{F}',
\label{eq:dsdt}
\ee
\be
{\delta s\over c_p}={1\over\alpha_T}\left({1\over\Gamma_1}{\delta p\over p}-{\delta\rho\over\rho}\right),
\label{eq:deltas}
\ee
where $s$, $T$, $c_p$, $\epsilon$, and $\pmb{F}$ are respectively the specific entropy,
temperature, specific heat at constant pressure, energy generation rate per gram, %{\bf 
and energy flux vector,%}
and %{\bf 
$\alpha_T=-\left({\partial\ln\rho/\partial\ln T}\right)_P$ is the volume expansion coefficient, $\nabla_{\rm ad}=\left({\partial\ln T/\partial\ln p}\right)_s$ is the adiabatic temperature gradient, %} 
and $\pmb{\xi}$ is the displacement vector, 
and $(')$ and $\delta$ indicate respectively the Eulerian and Lagrangian perturbations.
%{\bf 
Note that vectorial quantities are indicated by using italic bold faces. %}
%{\bf 
Equations (\ref{eq:eqmot}), (\ref{eq:cont}), (\ref{eq:dsdt}), and (\ref{eq:deltas}) are perturbed versions of
the equation of motion, continuity equation, entropy equation and the equation of state, respectively. %}
Note that we have applied the Cowling approximation, neglecting the Euler perturbation of the gravitational potential due to self-gravity, and that we have ignored the rotational deformation of the planet so that the equilibrium structure is spherical symmetric.

For the entropy equation (\ref{eq:dsdt}), we employ the approximation used by
Auclair-Desrotour \& Leconte (2018).
%{\bf 
Assuming the Newtonian cooling (e.g., Mihalas \& Mihalas 1999), %} 
the second term on the right-hand-side of equation (\ref{eq:dsdt}) may be approximately given by
\be
\nabla\cdot\pmb{F}'
\sim { \omega_D\rho c_p T}{T'\over T}
=\omega_D\rho c_p T\left({\delta s\over c_p}+\nabla_{\rm ad}{\delta p\over p}+\nabla V{\xi_r\over r}\right),
\label{eq:temppr}
\ee
where 
$V=-{d\ln p/ d\ln r}$, $\nabla={d\ln T/ d\ln p}$, 
\be
\omega_D={4\pi\over\tau_*}\left[\left({p\over p_*}\right)^{1/2}+\left({p\over p_*}\right)^2\right]^{-1},
\label{eq:omegad}
\ee
and $p_*$ is the pressure 
at the base of the heated layer, $\tau_*$ is the timescale parameter to specify the efficiency
of radiative cooling in the envelope, and we use $p_*=10^6{\rm dyn/cm^2}$ (see Auclair-Desrotour \& Leconte 2018; Iro et al 2005).
%{\bf 
Note that to derive equation (\ref{eq:temppr}) we have used the relations given by $\delta s/c_p=\delta T/T-\nabla_{\rm ad}\delta p/p$ and $\delta T/T=T'/T+\xi_rd\ln T/dr$. %}
%{\bf 
Substituting equation (\ref{eq:temppr}) into equation (\ref{eq:dsdt}), %} 
we obtain
the entropy perturbation $\delta s$ given by
\be
{\delta s\over c_p}={1\over\rmi\omega+\omega_D}{(\rho\epsilon)'\over \rho Tc_p}-{\omega_D\over\rmi\omega+\omega_D}\left(\nabla_{\rm ad}
{\delta p\over p}+\nabla V{\xi_r\over r}\right).
\ee

%{\bf 
In this paper, we assume uniform rotation for simplicity.
The convective core is likely to rotate uniformly if turbulent mixing is efficient enough in the core.
If this is the case, uniform rotation can be a good approximation since most of the mass of hot Jupiters is occupied by the convective core.
If there exists a strong differential rotation between the core and the envelope, the shear would excite turbulence, which
could weaken the differential rotation if the buoyant effects are weak (e.g., Turner 1979).
If there consistently exists a differential rotation between the core and the envelope, the frequency spectra of tidally excited $g$-modes and inertial
modes would be different from those found assuming uniform rotation since the frequency of inertial modes is simply proportional
to the rotation rate of the core but the frequency of $g$-modes is not. %}

\subsection{Forced Oscillation Equations}

In the presence of gravitational and/or thermal tidal forcing, the equations that govern the tidal responses
become a set of inhomogeneous linear differential equations.
Assuming the rotation axis of the planet is perpendicular to the orbital plane, 
the tidal potential in the planet due to the host star may be given by
\be
\Phi_T=-{GM_*\over |\pmb{r}-\pmb{a}_*(t)|}
=-GM_*\sum_{lm}W_{lm}{r^l\over a_*(t)^{l+1}}e^{-\rmi m\Phi}Y_l^m(\theta,\phi),
\ee
where the origin of spherical polar coordinates $(r,\theta,\phi)$ is at the centre of the planet, 
$\pmb{a}_*(t)$ is the position vector to the host star, $a_*(t)=|\pmb{a}_*(t)|$, $\Phi$ is the true anomaly, and
\be
W_{lm}=(-1)^{(l+m)/2}{\displaystyle\left[{4\pi\over 2l+1}(l-m)!(l+m)!\right]^{1/2}\over\displaystyle
\left[2^l\left({l-m\over 2}\right)!\left({l+m\over 2}\right)!\right]}=W_{l,-m},
\ee
which has a non-zero value for even values of $l-m$.
%{\bf 
Note that the tidal potential $\Phi_T$ does not contain the dipole terms associated with $l=1$. %}
Assuming that the eccentricity $e$ of the orbit is small so that $\Phi\approx \Omega_{\rm orb}t$, and taking only the dominant tidal component with $l=-m=2$, the tidal potential $\Phi_T$ in an inertial frame may be given by
\be
\Phi_T=-W_{2,-2}{GM_*\over a_*^3}r^2Y_2^{-2}(\theta,\phi)e^{2\rmi \Omega_{\rm orb}t}.
\label{eq:phit}
\ee
When the planet is uniformly rotating at $\Omega$, the tidal potential in the co-rotating frame may be given by
replacing $\phi$ by $\phi+\Omega t$ and hence the forcing frequency $2\Omega_{\rm orb}$ by 
$\omega=2\Omega_{\rm orb}+m\Omega=2(\Omega_{\rm orb}-\Omega)$ for $m=-2$.

Thermal tides are caused by insolation by the host star, which %{\bf 
produces the day and night sides on the
planet and %} 
is given by %{\bf 
(e.g., Auclair-Desrotour \& Leconte 2018) %}
\be
\epsilon'=\left\{\begin{array}{ll}
J_*(r,\theta,\phi,t)=\kappa_*F_*e^{-p(r)/p_*}\cos\phi_* & {\rm for} \quad 0\le\phi_*\le\pi/2\\
J_*(r,\theta,\phi,t)=0 & {\rm for} \quad \pi/2\le\phi_*\le\pi,
\end{array}\right.
\ee
where $\phi_*$ is the zenith angle of the host star as observed from the planet and is
given by $\cos\phi_*=\sin\theta\cos(\phi-\Phi)$, and
\be
F_*=\sigma_{\rm SB}T_*^4\left({R_*\over r_*}\right)^2, \quad \kappa_*={g_0\over p_*}, \quad g_0={GM\over R^2},
\ee
and $T_*$ and $R_*$ are respectively the surface temperature and radius of the host star, %{\bf 
$\kappa_*$ is the opacity at the base of the heated layer %} 
and $r_*$ is the distance between the planet and the host star, set equal to the semi-major axis $a_*$ of the orbit.
Assuming $\Phi=\Omega_{\rm orb}t$,
we obtain
\be
\int_0^\pi d\theta \sin\theta\int_{\Phi-\pi/2}^{\Phi+\pi/2} d\phi  (Y_2^{-2})^*\cos\phi_*={1\over 16}\sqrt{15\pi\over 2}e^{2\rmi\Omega_{\rm orb}t}.
\ee
In general, in the co-rotating frame of the planet, we may expand $J_*$ in terms of spherical harmonic function as
\be
J_*=\sum_{l,m,n}J^{(lmn)}_*(r)Y_l^m(\theta,\phi)e^{\rmi\omega_{mn} t},
\ee
where
$\omega_{mn}=n\Omega_{\rm orb}+m\Omega$ and $n\Omega_{\rm orb}$ represents the forcing frequency in an
inertial frame.
If we take only the component with $n=-m=2$, for which $\omega_{-2,2}=2(\Omega_{\rm orb}-\Omega)$, 
assuming $(\rho\epsilon)'=\rho\epsilon'$ %{\bf 
since the unperturbed state has no insolation so that $\epsilon=0$, %} 
we obtain
\be
{\delta s\over c_p}={1\over\rmi\omega+\omega_D}{\epsilon'\over Tc_p}-{\omega_D\over\rmi\omega+\omega_D}\left(\nabla_{\rm ad}
{\delta p\over p}+\nabla V{\xi_r\over r}\right),
\label{eq:dsocp}
\ee
where $\omega=\omega_{-2,2}$ and
\be
\epsilon'={1\over 16}\sqrt{15\pi\over 2}\kappa_*F_*e^{-p(r)/p_*}Y_2^{-2}(\theta,\phi)e^{\rmi\omega t}.
\ee

To represent the perturbations of tidally perturbed and rotating planets, we employ series expansion in terms of spherical harmonic functions
$Y_l^m(\theta,\phi)$ for a given $m$ with different $l$s (e.g., Lee \& Saio 1986).
The pressure perturbation is given by
\be
p'(r,\theta,\phi,t)=\sum_lp'_l(r)Y_l^m(\theta,\phi)e^{\rmi\omega t}, 
\label{eq:prhoperturbations}
\ee
and the displacement vector $\pmb{\xi}$ by
\be
\xi_r(r,\theta,\phi,t)=r\sum_{l}S_l(r)Y_l^m(\theta,\phi)e^{\rmi\omega t},
\ee
\be
\xi_\theta(r,\theta,\phi,t)=r\sum_{l,l'}\left[H_l(r){\partial\over\partial\theta}Y_l^m(\theta,\phi)+T_{l'}{1\over\sin\theta}{\partial\over\partial\phi}Y_{l'}^m(\theta,\phi)\right]e^{\rmi\omega t},
\ee
\be
\xi_\phi(r,\theta,\phi,t)=r\sum_{l,l'}\left[H_l(r){1\over\sin\theta}{\partial\over\partial\phi}Y_l^m(\theta,\phi)-T_{l'}{\partial\over\partial\theta}Y_{l'}^m(\theta,\phi)\right]e^{\rmi\omega t},
\label{eq:disp_phi}
\ee
where $l_j=|m|+2(j-1)$ and $l'_j=l_j+1$ for even modes, and $l_j=|m|+2j-1$ and $l'_j=l_j-1$ for odd modes
and $j=1,~2,~\cdots,~j_{\rm max}$.
In this paper, since we assume for simplicity that the spin axis of the planet is perpendicular to the orbital plane of the planet and that the planet is on a circular orbit around the host star,
we retain only the forcing terms proportional to $Y_2^{-2}e^{\rmi\omega t}$ with $\omega=2(\Omega_{\rm orb}-\Omega)$, for which the tidal responses may be represented by the sum of terms proportional to
$Y_l^{-2}e^{\rmi\omega t}$ with $l=2j$ and $j=1,~2, ~\cdots,~j_{\rm max}$.
Note that $j_{\rm max}$ determines the length of the series expansions for the responses, and we use $j_{\rm max}=12$
in this paper.

Substituting the expansions into the perturbed basic equations, we obtain a set of 
linear ordinary differential equations with inhomogeneous terms.
Defining the dependent variables as
\be
\pmb{y}_1=(S_{l_j}), \quad \pmb{y}_2=\left({p'_{l_j}\over\rho g r}\right), 
\quad \pmb{y}_6=\left({\delta s_{l_j}\over c_p}\right), \quad \pmb{h}=(H_{l_j}), 
\quad \pmb{t}=(T_{l'_j}), \quad \pmb{Y}_2=\pmb{y}_2+{\pmb{\psi}\over gr},
\label{eq:depvar}
\ee
the perturbed basic equations reduce to
\be
r{d\pmb{y}_1\over dr}=\left({V\over\Gamma_1}-3\right)\pmb{y}_1-{V\over\Gamma_1}\pmb{Y}_2+\pmbmt{\Lambda}_0\pmb{h}+\alpha_T\pmb{y}_6+{V\over\Gamma_1}{\pmb{\psi}\over gr},
\ee
\be
r{d\pmb{Y}_2\over dr}=(c_1\bar\omega^2+rA)\pmb{y}_1+(1-U-rA)\pmb{Y}_2-2c_1\bar\omega\bar\Omega\left(m\pmb{h}+\pmbmt{C}_0\rmi\pmb{t}\right)
+\alpha_T\pmb{y}_6+rA{\pmb{\psi}\over gr},
\ee
\be
-\pmbmt{M}_0\pmb{h}+\pmbmt{L}_1\rmi\pmb{t}=-\nu\pmbmt{K}\pmb{y}_1,
\label{eq:auxiliary01}
\ee
\be
\pmbmt{L}_0\pmb{h}-\pmbmt{M}_1\rmi\pmb{t}=m\nu\pmbmt{\Lambda}_0^{-1}\pmb{y}_1+{1\over c_1\bar\omega^2}\pmb{Y}_2,
\label{eq:auxiliary02}
\ee
\be
\pmb{y}_6
={1\over\rmi\omega+\omega_D}\pmb{j}_*-{\omega_D\over\rmi\omega+\omega_D} V
\left[\nabla_{\rm ad}\pmb{Y}_2-\nabla_{\rm ad}{\pmb{\psi}\over gr}+\left(\nabla-\nabla_{\rm ad}\right)\pmb{y}_1\right],
\label{eq:y6}
\ee
where $U={d\ln M_r/ d\ln r}$, $c_1={(r/R_p)^3/ (M_r/M)}$, and
\be
\nu={2\Omega\over\omega}, \quad \bar\omega={\omega\over\sigma_0}, \quad \bar\Omega={\Omega\over\sigma_0}, \quad \sigma_0=\sqrt{GM\over R^3},
\ee
and non-zero elements of the matrices $\pmbmt{\Lambda}_0$, $\pmbmt{C}_0$, $\pmbmt{L}_0$, $\pmbmt{L}_1$, $\pmbmt{K}$, $\pmbmt{M}_0$, $\pmbmt{M}_1$ for even modes
are defined by
\be
(\pmbmt{\Lambda}_0)_{j,j}=l_j(l_j+1), \quad
(\pmbmt{C}_0)_{j,j}=-(l_j+2)J^m_{l_j+1}, \quad (\pmbmt{C}_0)_{j+1,j}=(l_j+1)J^m_{l_j+2}
\ee
\be
(\pmbmt{L}_0)_{j,j}=1-{m\nu\over l_j(l_j+1)}, \quad (\pmbmt{L}_1)_{j,j}=1-{m\nu\over l'_j(l'_j+1)}, \quad (\pmbmt{K})_{j,j}={J^m_{l_j+1}\over l_j+1}, 
\quad (\pmbmt{K})_{j,j+1}=-{J^m_{l_j+2}\over l_j+2},
\ee
\be
(\pmbmt{M}_0)_{j,j}=\nu{l_j\over l_j+1}J^m_{l_j+1}, 
\quad (\pmbmt{M}_0)_{j,j+1}=\nu{l_j+3\over l_j+2}J^m_{l_j+2}, \quad 
(\pmbmt{M}_1)_{j,j}=\nu{l_j+2\over l_j+1}J^m_{l_j+1}, \quad 
(\pmbmt{M}_1)_{j+1,j}=\nu{l_j+1\over l_j+2}J^m_{l_j+2}.
\ee
%{\bf 
See the Appendix A for the derivation of equations (30) to (34). %}
The vectors $\pmb{j}_*$ and $\pmb{\psi}$ are inhomogeneous forcing terms and have only the first component given respectively by
\be
(\pmb{j}_*)_1={\sqrt{15\pi/2}\over 16}{\kappa_*F_*\over Tc_p}e^{-p(r)/p_*},
\ee
and
\be
{(\pmb{\psi})_1\over gr}={\Phi_T\over gr}=-\sqrt{3\pi\over 10}{M_*\over M_*+M}c_1{\bar\Omega^2_{\rm orb}} 
\ee
with $\bar\Omega_{\rm orb}={\Omega_{\rm orb}/\sigma_0}$.
To estimate the temperature $T$ and the specific heat $c_p$, we assume those for an ideal gas, that is,
\be
T={\mu\over{\cal R}}{p\over\rho}, \quad c_p={5\over 2}{{\cal R}\over\mu},
\ee
and $\cal R$ is the gas constant, and $\mu$ is the mean molecular weight, for which we use $\mu=1.3$.

Using the auxiliary equations (\ref{eq:auxiliary01}) and (\ref{eq:auxiliary02}), we obtain
\be
\pmbmt{\Lambda}_0\pmb{h}={\pmbmt{W}\over c_1\bar\omega^2}\pmb{Y}_2+\nu\pmbmt{W}\pmbmt{O}\pmb{y}_1,
\label{eq:auxiliary21}
\ee
\be
2c_1\bar\omega\bar\Omega(m\pmb{h}+\pmbmt{C}_0\rmi\pmb{t})=\nu\pmbmt{O}^T\pmbmt{W}\pmb{Y}_2+4c_1\bar\Omega^2\pmbmt{G}\pmb{y}_1,
\label{eq:auxiliary22}
\ee
where
\be
\pmbmt{W}=\pmbmt{\Lambda}_0(\pmbmt{L}_0-\pmbmt{M}_1\pmbmt{L}_1^{-1}\pmbmt{M}_0)^{-1}, \quad 
\pmbmt{O}=m\pmbmt{\Lambda}_0^{-1}-\pmbmt{M}_1\pmbmt{L}_1^{-1}\pmbmt{K}, \quad 
\pmbmt{G}=\pmbmt{O}^T\pmbmt{W}\pmbmt{O}-\pmbmt{C}_0\pmbmt{L}_1^{-1}\pmbmt{K}.
\ee 
Substituting equations (\ref{eq:auxiliary21}), (\ref{eq:auxiliary22}), and (\ref{eq:y6}), we obtain
the set of linear differential equations for forced oscillations:
\be
r{d\pmb{y}_1\over dr}
=\left[\left({V\over\Gamma_1}-3+\beta_1\right)\pmbmt{1}+\nu\pmbmt{W}\pmbmt{O}\right]\pmb{y}_1
+\left[{\pmbmt{W}\over c_1\bar\omega^2}-\left({V\over\Gamma_1}+\beta_2\right)\pmbmt{1}\right]\pmb{Y}_2
+{\alpha_T\over\rmi\omega+\omega_D}\pmb{j}_*+\left({V\over\Gamma_1}+\beta_2\right){\pmb{\psi}\over gr},
\label{eq:dy1}
\ee
\be
r{d\pmb{Y}_2\over dr}
=\left[\left(c_1\bar\omega^2+rA+\beta_1\right)\pmbmt{1}-4c_1\bar\Omega^2\pmbmt{G}\right]\pmb{y}_1
+\left[\left(1-U-rA-\beta_2\right)\pmbmt{1}-\nu\pmbmt{O}^T\pmbmt{W}\right]\pmb{Y}_2
+{\alpha_T\over\rmi\omega+\omega_D}\pmb{j}_*+\left(rA+\beta_2\right){\pmb{\psi}\over gr},
\label{eq:dy2}
\ee
where $\pmbmt{1}$ is the unit matrix, and 
\be
\beta_1={\alpha_T(\nabla_{\rm ad}-\nabla)V\over\rmi\omega/\omega_D+1}, \quad
\beta_2={\alpha_T\nabla_{\rm ad}V\over\rmi\omega/\omega_D+1}.
\ee
The boundary condition at the centre is the regularity condition of the functions $\pmb{y}_1$
and $\pmb{Y}_2$ {\bf }(see the Appendix B).
The outer boundary condition at the surface of the planet is given by $\delta p=0$
%{\bf 
(see, e.g., Unno et al 1989). %}

\section{Numerical Results}

The tidal torque on the planet may be given by %{\bf 
(e.g., Auclair-Desrotour \& Leconte 2018) %}
\be
{\cal N}=-\int\overline{{\partial\Phi_T\over\partial\phi}\rho'}dV
=-{1\over 2}\int{\rm Re}\left({\partial\Phi_T\over\partial\phi}\rho'^*\right)dV
=-\int{\rm Im}\left({\Phi_T}\rho'^*\right)dV,
\label{eq:tidaltorque00}
\ee
where $\rho'$ is the tidal response caused by the tides associated with $\Phi_T$ and/or $\epsilon'$.
If the density perturbation in the rotating planet is represented by the series expansion 
similar to (\ref{eq:prhoperturbations}) and the tidal potential $\Phi_T(r,\theta,\phi)$ is simply proportional to $Y_2^{-2}(\theta,\phi)$ %{\bf 
as given by equation (\ref{eq:phit}), %}
the tidal torque $\cal N$ is computed by
\be
{\cal N}=\sqrt{3\pi\over 10}{GM_*\over a_*^3}\int_0^Rdrr^4{\rm Im}[\rho'^*_2(r)],
\label{eq:tidaltorque}
\ee
where the tidal response $\rho'_2(r)$ is obtained by solving the inhomogeneous linear differential equations
(\ref{eq:dy1}) and (\ref{eq:dy2}) for a given forcing frequency $\omega$.
For numerical computations, we assume $M_*=M_\odot$, $R_*=R_\odot$, and $T_*=5.8\times10^3$K for the host star, and the distance between the planet and the star is assumed to be $r_*=a_*=0.05$A.U..

\begin{figure}
\resizebox{0.45\columnwidth}{!}{
\includegraphics{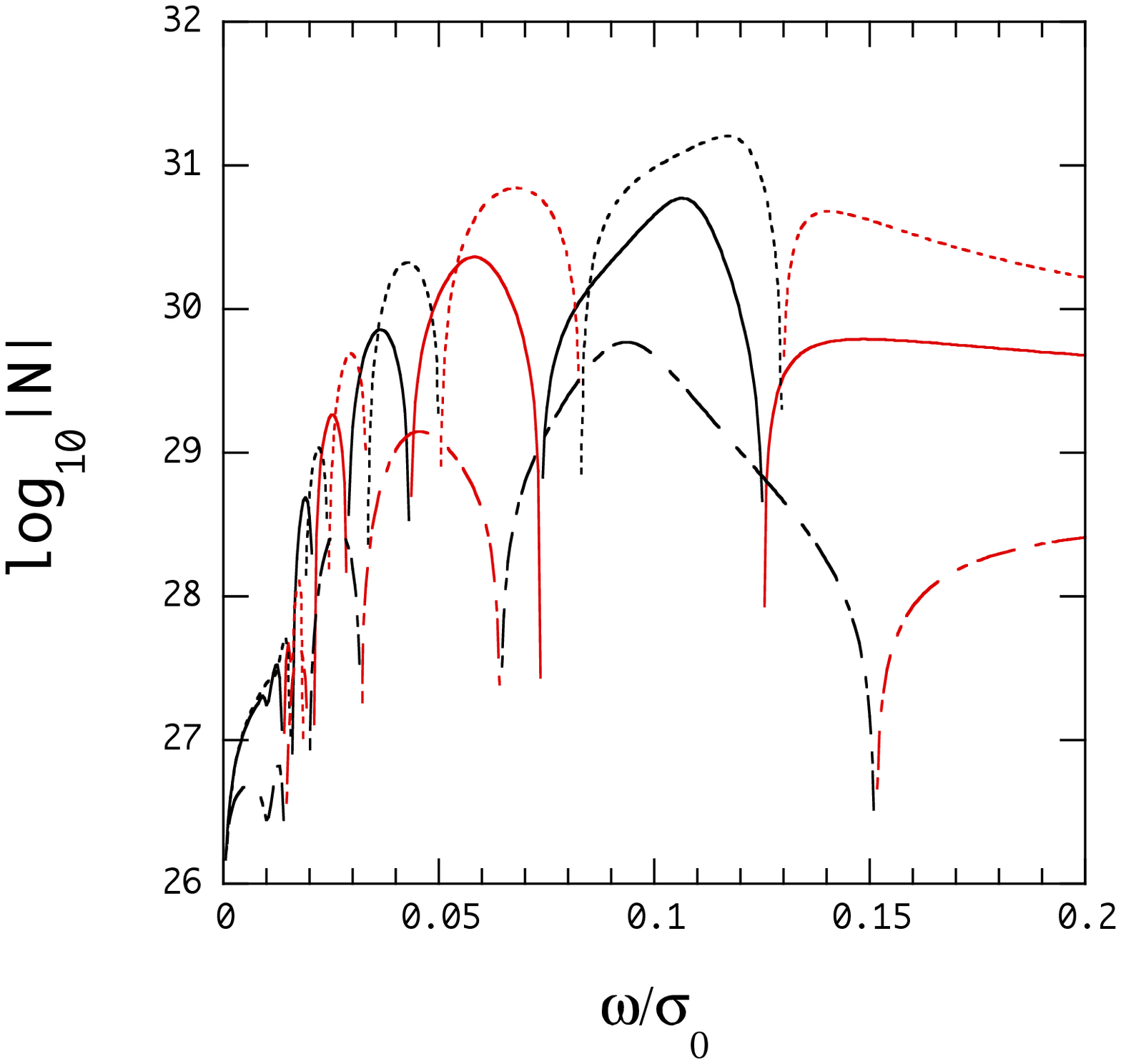}}
\hspace*{0.5cm}
\resizebox{0.45\columnwidth}{!}{
\includegraphics{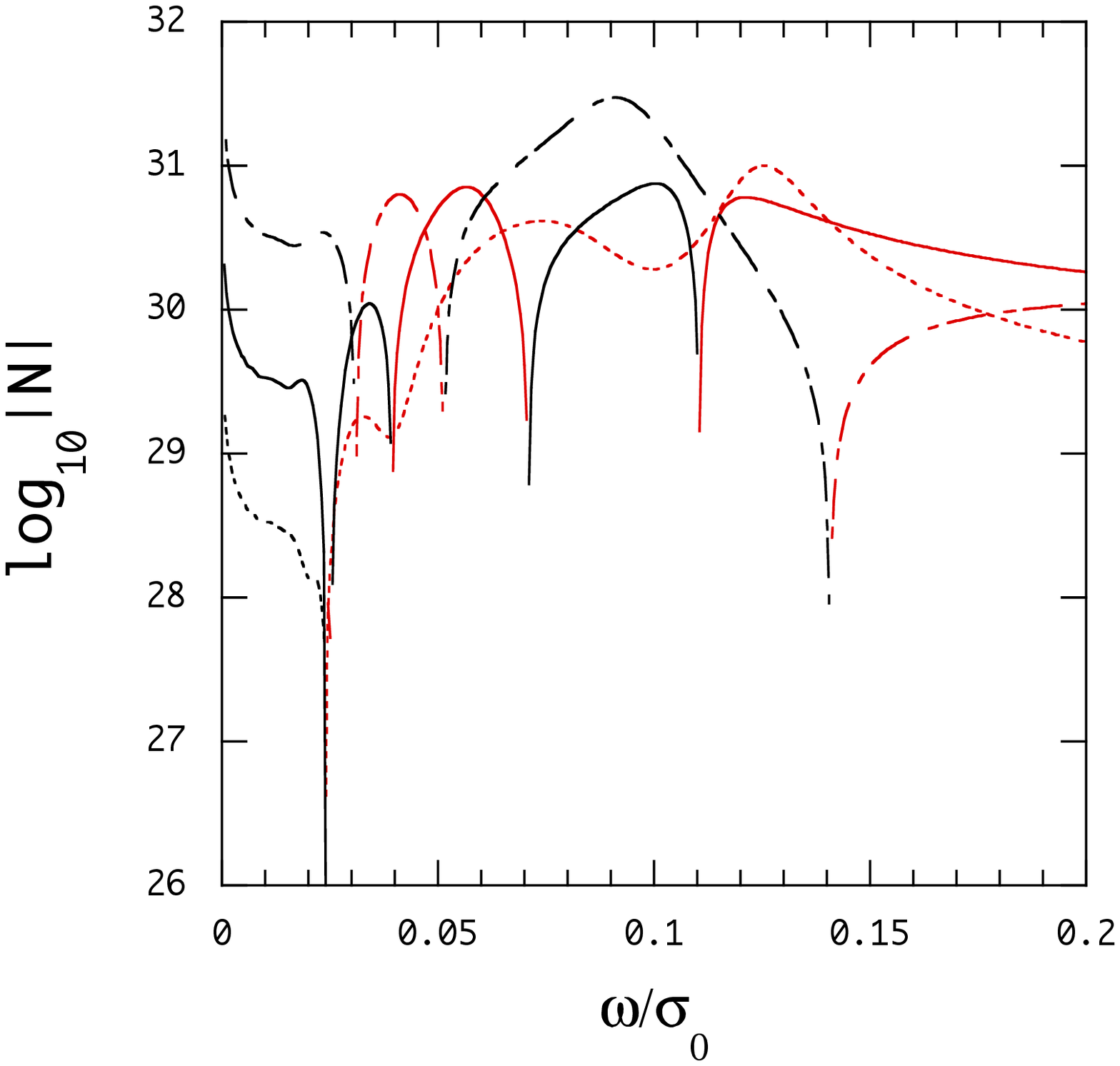}}
\caption{Tidal torque, given in erg, due to thermal tides (left panel) and gravitational tides (right panel)
as a function of the forcing frequency $\bar\omega=\omega/\sigma_0$
for $\bar\Omega=0$, where the red (black) lines are for positive
(negative) $\cal N$, and the dash-dotted lines, solid lines, and dotted lines are
for $\tau_*=0.1$, 1, and 10 days, respectively.
}
\end{figure}

We first discuss the case of $\bar\Omega=0$.
We calculate non-adiabatic free $g$-modes propagating in the radiative envelope of the planet
for different values of $\tau_*$, which determines the thermal time scales in the envelope.
Free $g$ modes may be computed by setting $\pmb{\psi}=0$ and $\pmb{j}_*=0$  
in equations (\ref{eq:dy1}) and (\ref{eq:dy2}) and
introducing a normalization condition, for example, given by $S_{l_1}=1$ at the surface.
Note that non-adiabatic $g$-modes have complex eigenfrequency $\bar\omega$.
The result of non-adiabatic calculation of free $g$-modes is summarized in Table 1, in which
complex eigenfrequency $\bar\omega$ is given for several low radial order $g$ modes for
three values of the parameter $\tau_*$.
The shorter $\tau_*$ is, the larger the non-adiabatic effects are.
The table shows that the $g$-modes are all pulsationally stable and have large damping rate 
$\eta\equiv\omega_{\rm I}/\omega_{\rm R}\gtsim 0.1$, %{\bf 
where $\omega_{\rm R}$ and $\omega_{\rm I}$
are the real and imaginary parts of the complex frequency $\omega$ (see the caption to Table 1). %}
It also shows that the frequency $\bar\omega_{{\rm R},n}$ of the $g_n$-mode increases
but the frequency difference  
$\bar\omega_{{\rm R},n}-\bar\omega_{{\rm R},n+1}$ decreases as the radial order $n$ and $\tau_*$ increase.

\begin{table*}
\begin{center}
\caption{Complex eigenfrequency $\bar\omega=\bar\omega_{\rm R}+\rmi\bar\omega_{\rm I}$ of $l=2$ $g_n$-modes in the envelope for $\bar\Omega=0$.}
\begin{tabular}{@{}ccccccc}
\hline
$\tau_*$ (day)  & \hspace*{2cm}0.1 && \hspace*{2cm}1 && \hspace*{2cm}10 & \\
\hline
 $n$ & $\bar\omega_{\rm R}$ & $\bar\omega_{\rm I}$ & $\bar\omega_{\rm R}$ & $\bar\omega_{\rm I}$ & $\bar\omega_{\rm R}$ & $\bar\omega_{\rm I}$  \\
\hline
1 & $9.29\times10^{-2}$ & $1.36\times10^{-2}$ & $1.09\times10^{-1}$ & $1.04\times10^{-2}$ & $1.24\times10^{-1}$ & $1.06\times10^{-2}$\\
2 & $4.15\times10^{-2}$ & $1.24\times10^{-2}$ & $5.86\times10^{-2}$ & $1.24\times10^{-2}$ & $7.98\times10^{-2}$ & $1.97\times10^{-2}$\\
3 & $2.40\times10^{-2}$ & $8.50\times10^{-3}$ & $3.62\times10^{-2}$ & $9.63\times10^{-3}$ & $5.77\times10^{-2}$ & $1.96\times10^{-2}$\\
4 & $1.64\times10^{-2}$ & $6.33\times10^{-3}$ & $2.54\times10^{-2}$ & $7.52\times10^{-3}$ & $4.25\times10^{-2}$ & $1.38\times10^{-2}$\\
\hline
\end{tabular}
\medskip
\end{center}
\end{table*}

Figure 2 plots the absolute value of the tidal torque ${\cal N}$ for $\bar\Omega=0$ as a function
of the forcing frequency $\bar\omega$ for three different values of $\tau_*$, where
the left panel is for the case of $\pmb{\psi}=0$ and $\pmb{j}_*\not=0$, and the right panel is for 
the case of $\pmb{\psi}\not=0$ and $\pmb{j}_*=0$ in equations (\ref{eq:dy1}) and (\ref{eq:dy2}).
For both cases, there appears, as a function of $\bar\omega$, broad peaks of 
$|{\cal N}|$, which are
produced by frequency resonance between the forcing frequency and the natural frequency of the $g$-modes.
The width of the peaks may be determined by the magnitude of $\tau_*$, that is, the larger $\tau_*$ is,
the narrower the peaks are, which is partly because the frequency difference  
$\bar\omega_{{\rm R},n}-\bar\omega_{{\rm R},n+1}$ decreases with increasing $\tau_*$ for the $g$-modes
with large damping rates $\eta$.

The sign of the tidal torque at the resonance peaks alternately changes as the $g$-mode in resonance with
the forcing is changed with decreasing $\bar\omega_{\rm R}$, except for the case of
the gravitational tide for $\tau_*=10$day.
The response to the tidal potential $\Phi_T$ is quite similar to that to the thermal tides
except for very low tidal frequency region.
The tidal torque $|{\cal N}|$ due to the thermal tides decreases as $\bar\omega\rightarrow0$, but
the torque due to the gravitational tides
tend to a constant value, corresponding to that
of the gravitational equilibrium tide, the magnitudes of which is proportional to $\tau_*^{-1}$, that is, the possible amount of energy dissipation in the envelope.
Since we consider no effects of turbulent fluid motion in the convective core on tidal
responses due to the gravitational tidal potential $\Phi_T$, we have to be cautious about estimating the
tidal effects on the planets.

\begin{figure}
\resizebox{0.45\columnwidth}{!}{
\includegraphics{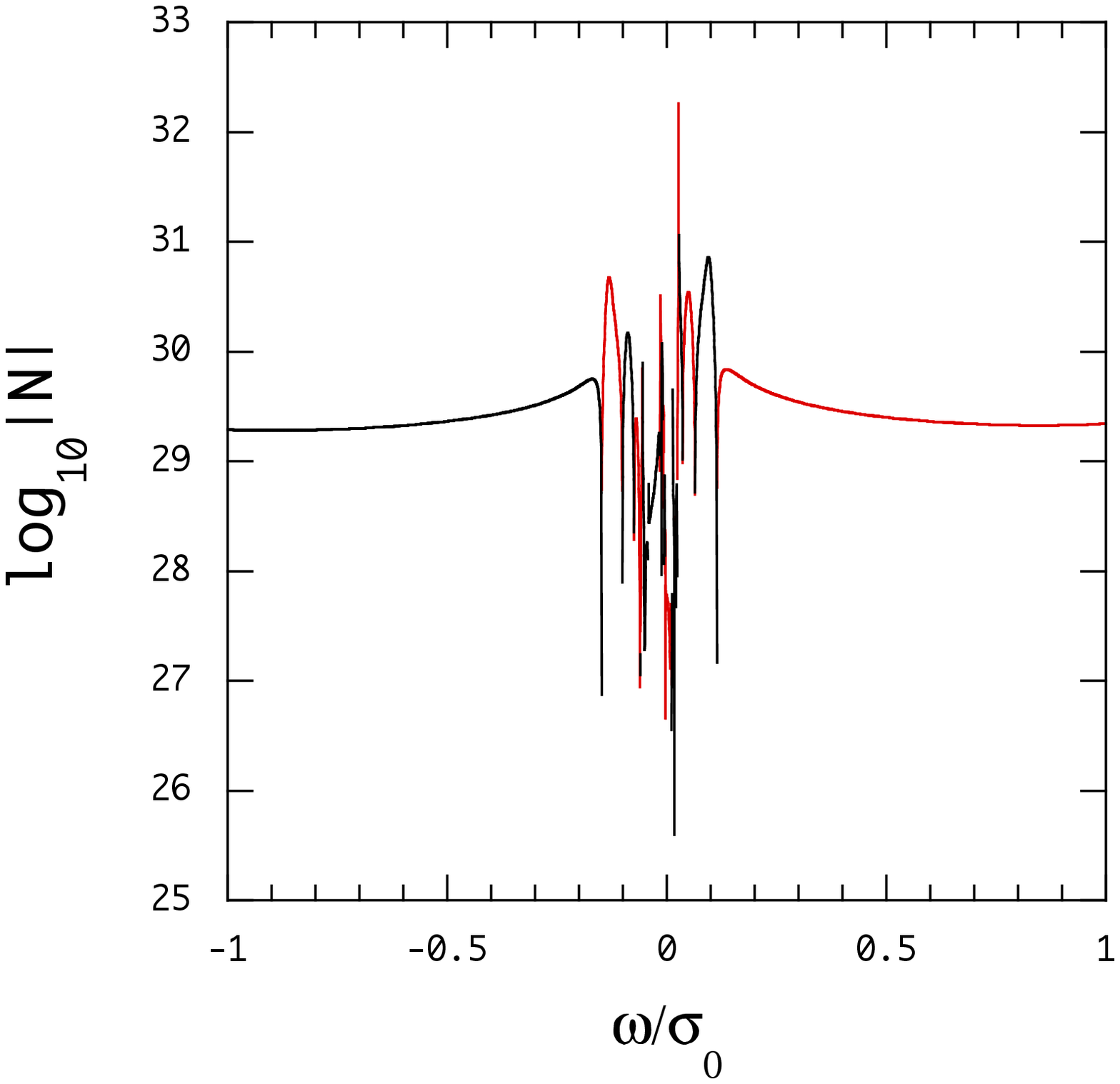}}
\hspace*{0.5cm}
\resizebox{0.45\columnwidth}{!}{
\includegraphics{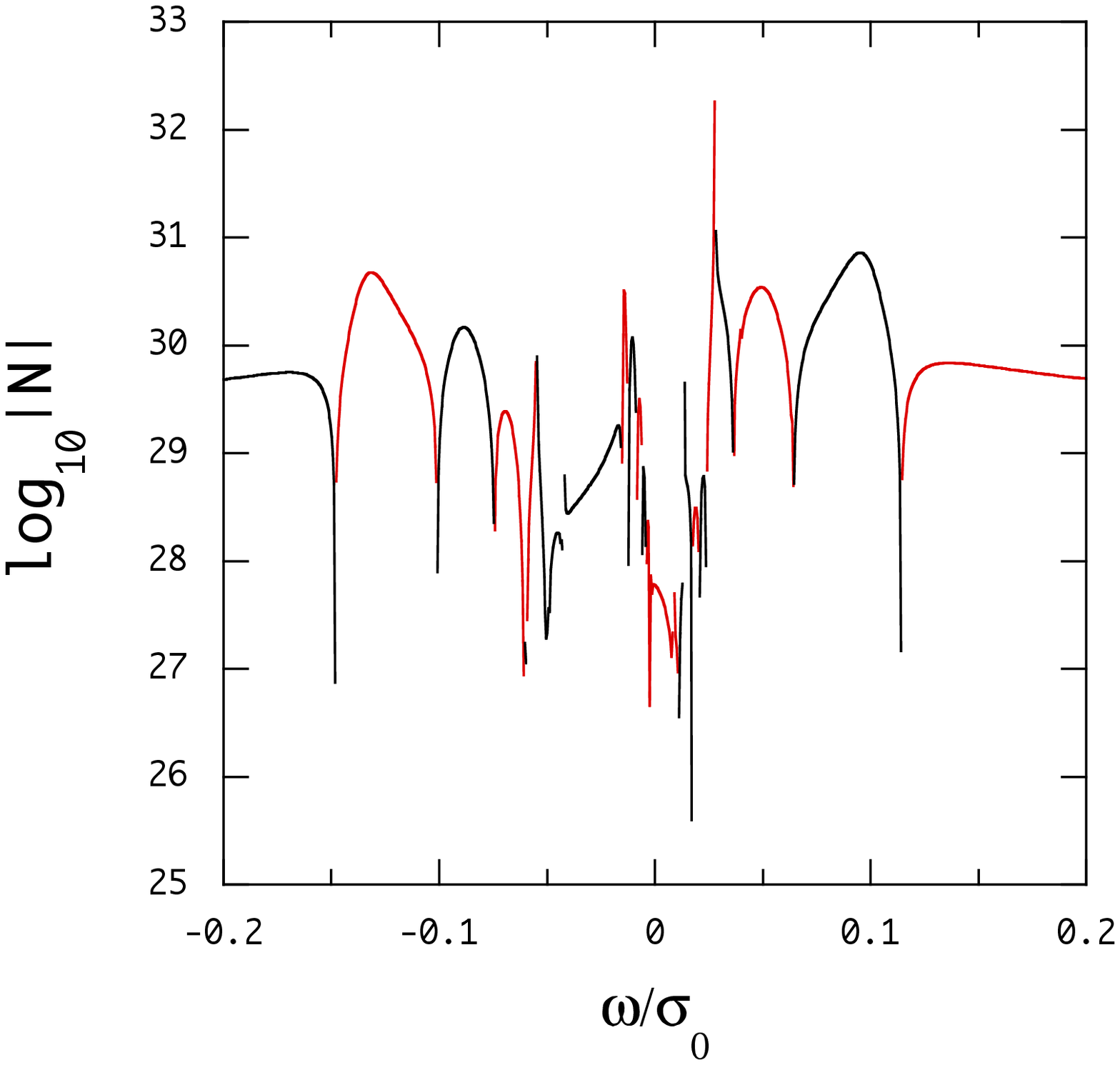}}
\caption{Tidal torque, given in erg, due to thermal tides ($\pmb{j}_*\not=0$ and $\pmb{\psi}=0$) as a function of the forcing frequency $\bar\omega$
for $\bar\Omega=0.05$, where positive (negative)
$\bar\omega$ indicates prograde (retrograde) forcing, and the red (black) lines are for positive
(negative) $\cal N$.
}
\end{figure}

\begin{figure}
\resizebox{0.45\columnwidth}{!}{
\includegraphics{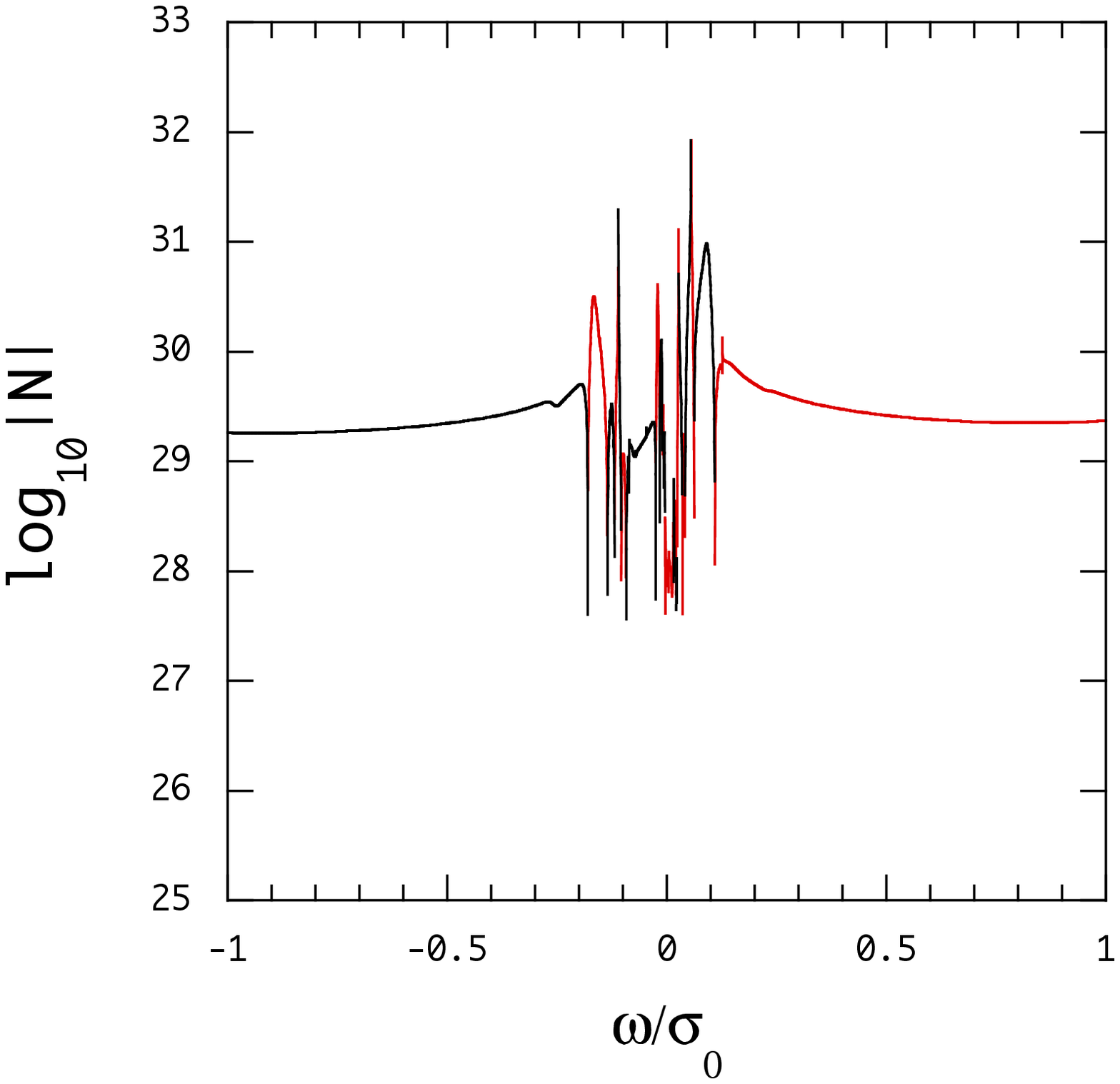}}
\hspace*{0.5cm}
\resizebox{0.45\columnwidth}{!}{
\includegraphics{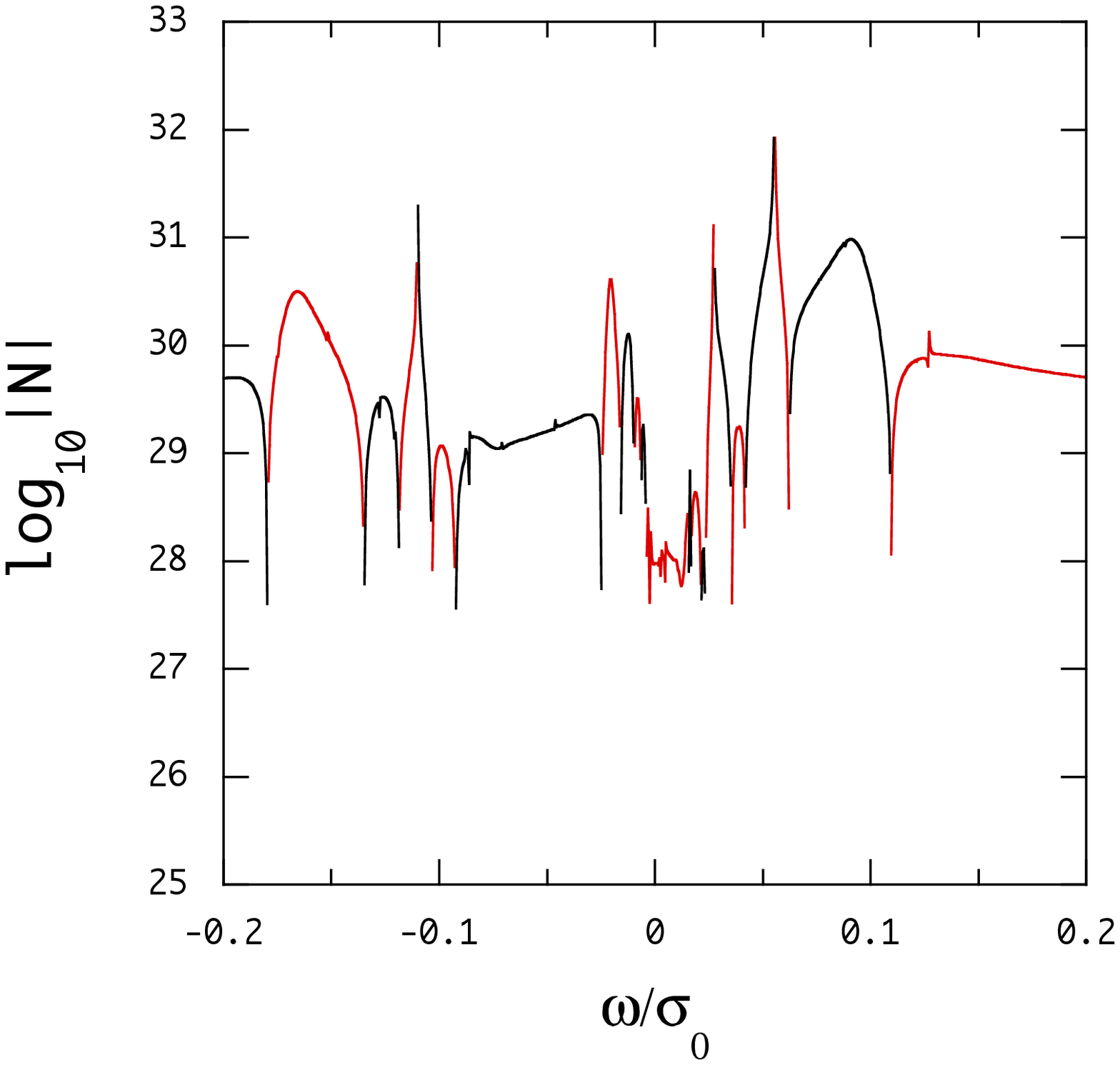}}
\caption{Same as Figure 3 but for $\bar\Omega=0.1$.
}
\end{figure}

\begin{table*}
\begin{center}
\caption{Complex eigenfrequency $\bar\omega=\bar\omega_{\rm R}+\rmi\bar\omega_{\rm I}$ of $g_n$-modes, $r_n$-modes 
and inertial modes $i_L$ with $L\equiv l_0-|m|$ for $m=-2$ and $\bar\Omega=0.1$, where we use $\tau_*=$1 day.}
\begin{tabular}{@{}ccccc}
\hline
  & \hspace*{2cm}prograde && \hspace*{2cm}retrograde & \\
\hline
 modes & $\bar\omega_{\rm R}$ & $\bar\omega_{\rm I}$ & $\bar\omega_{\rm R}$ & $\bar\omega_{\rm I}$   \\
\hline
$g_1$ & $9.36\times10^{-2}$ & $9.47\times10^{-3}$ & $-1.68\times10^{-1}$ & $9.08\times10^{-3}$ \\
$g_2$ & $4.85\times10^{-2}$ & $1.07\times10^{-2}$ & $-1.23\times10^{-1}$ & $1.26\times10^{-2}$ \\
$g_3$ & $2.95\times10^{-2}$ & $8.05\times10^{-3}$ & $-9.91\times10^{-2}$ & $1.29\times10^{-2}$ \\
\hline
$r_1$ & $\cdots$ & $\cdots$ & $-2.07\times10^{-2}$ & $1.60\times10^{-3}$ \\
$r_2$ & $\cdots$ & $\cdots$ & $-1.19\times10^{-2}$ & $2.53\times10^{-3}$ \\
$r_3$ & $\cdots$ & $\cdots$ & $-7.47\times10^{-3}$ & $2.05\times10^{-3}$ \\
\hline
$i_2$ & $5.55\times10^{-2}$ & $1.16\times10^{-5}$ & $-1.10\times10^{-1}$ & $1.49\times10^{-6}$ \\
$i_4$ & $1.27\times10^{-1}$ & $9.89\times10^{-6}$ & $-1.52\times10^{-1}$ & $1.33\times10^{-6}$ \\
$i_4$ & $2.74\times10^{-2}$ & $3.35\times10^{-6}$ & $-8.60\times10^{-2}$ & $4.06\times10^{-6}$ \\
\hline
\end{tabular}
\medskip
\end{center}
\end{table*}

Let us briefly discuss the modal properties of low frequency, even parity, free oscillation modes of the rotating planets, such as $g$-modes, $r$-modes and inertial modes.
Oscillation modes of rotating planets are separated into prograde modes and retrograde modes, observed in
the co-rotating frame of the planet.
In our convention, for negative $m$, prograde (retrograde) modes correspond to positive (negative) spin parameter $\nu=2\Omega/\omega$ where $\omega$ is the oscillation frequency observed in the co-rotating frame of the planet.
The modal properties of low frequency $g$-modes, frequency and stability, are affected by rotation, particularly when $|\nu|\gtsim 1$.
Rotation also produces new kinds of oscillation modes, called inertial modes and $r$-modes.
Note that inertial modes propagate in nearly isentropic regions and that $r$-modes, which form
a subclass of inertial modes, appear only as retrograde modes.
The restoring force for inertial modes is the Coriolis force, and
their frequency $\omega$ is proportional to the rotation frequency $\Omega$ and 
the ratio $\omega/\Omega$ is limited to $|\omega/\Omega|\le 2$.
In other wards, inertial modes appear when $|\nu|\ge 1$.
The radiative envelope is the propagation regions of $r$-modes of even parity, for which both
Coriolis force and buoyant force plays essential roles.
It is well known that the asymptotic frequency $\omega$ of $r$-modes in the limit of $\Omega\rightarrow 0$ is given by $2m\Omega/[l'(l'+1)]$.
For $m=-2$, we compute free non-adiabatic $g$-modes, $r$-modes, and inertial modes of the planet model
for $\bar\Omega=0.1$ and tabulate their complex eigenfrequency in Table 2, where we have assumed $\tau_*=1$day.
Oscillation frequency $\bar\omega$ of the low radial order $g$-modes in the envelope shows differences between prograde and retrograde modes for $\bar\Omega=0.1$ and $|\bar\omega|\ltsim0.1$.
For even parity $r$-modes of $m=-2$, the frequencies $\bar\omega$ tabulated in Table 2 are
significantly different from the asymptotic value $2m\bar\Omega/[l'(l'+1)]$, 
which is $-1/3$ for $\bar\Omega=0.1$.
The inertial modes belonging to $L\equiv l_0-|m|=2$ and 4 are tabulated in Table 2.
For the classification using $l_0-|m|$, see, e.g., Yoshida \& Lee (2000), who computed
inertial modes of isentropic polytropes to tabulate $\kappa_0=\omega/\Omega$
for different values of $m$ and the polytropic index $n$, where $\kappa_0$ was estimated in the limit of
$\Omega\rightarrow 0$.
The ratio $\omega/\Omega$ for the inertial modes in Table 2 in this paper is consistent
with the value of $\kappa_0$ computed for the $m=2$ inertial modes 
of the $n=1$ polytrope, see Table 1 of Yoshida \& Lee (2000).
Note that for positive $m$, prograde (retrograde) modes have negative (positive) $\kappa_0$ for $\bar\Omega>0$.
Since the inertial modes are confined in the convective core where non-adiabatic effects are negligible,
the imaginary part of the inertial mode frequency is much smaller than that of the $g$-modes and $r$-modes, which are confined in 
the radiative envelope where non-adiabatic effects are very large.

Assuming only the thermal tides operate (i.e., $\pmb{j}_*\not=0$ and $\pmb{\psi}=0$),
we compute the tidal torque $\cal N$ as a function of the forcing frequency $\bar\omega$
for a fixed value of $\bar\Omega$ for $\tau_*=1$day.
The plots of $|\cal N|$ for $\bar\Omega=0.05$ and 0.1 are respectively given in Figures 3 and 4, where
positive and negative $\bar\omega$ corresponds
to prograde and retrograde forcing observed in the co-rotating frame of the planet, 
and the red (black) lines represent positive (negative) parts of $\cal N$.
The left panels show the torque in the range of $|\bar\omega|\le 1$ and the right panels for $|\bar\omega|\le 0.2$
as a magnification.
The tidal torque only weakly depends on $\bar\omega$ for $|\bar\omega|\gtsim 0.2$.
However, there appear broad and sharp peaks of $|\cal N|$ in the range of $|\bar\omega|\ltsim 0.2$.
Comparing the frequency $\bar\omega_P$ at the peaks with the natural frequency $\bar\omega_{\rm R}$ of the low frequency modes tabulated in Table 2,
we find that the broad peaks are produced when the forcing frequency is in resonance with
the natural frequency of the $g$-modes and $r$-modes in the envelope, and that the sharp peaks are produced by the resonance with the inertial modes in the core.
The width of the peaks may reflect the magnitude of $\bar\omega_{\rm I}$ of the modes
in resonance with the forcing, that is,
if the modes in resonance have $|\bar\omega_{\rm I}|\sim|\bar\omega_{\rm R}|$, the peaks will be broad, 
while if they have $|\bar\omega_{\rm I}|\ll|\bar\omega_{\rm R}|$ the peaks will be very sharp.
Comparing the two cases of $\bar\Omega=0.05$ and 0.1, the frequency $\bar\omega_P$ of the broad peaks due to the $g$-modes does not significantly depend on $\bar\Omega$, which is particularly the case for the peaks on the prograde sides.
It is also interesting to note that the frequency $\bar\omega_P$ of the peaks due to
the $r$-modes does not show strong dependence on $\bar\Omega$, which is because 
the frequency of $r$-modes propagating in a geometrically thin atmosphere becomes insensitive to
the rotation speed $\Omega$ for rapid rotation (see, e.g., Pedlosky 1986).
The peak frequency $\bar\omega_P$ due to the inertial modes, however, linearly depends on the rotation
frequency $\bar\Omega$ since the natural frequency $\omega\propto\Omega$ for inertial modes.
For example, for $\bar\Omega=0.1$, the sharp peaks located at the frequency $\bar\omega\approx 0.055$ and $-0.11$ respectively correspond to the inertial modes with the ratio
$\omega/\Omega\approx 0.55$ and $-1.1$ belonging to $l_0-|m|=2$ (see Yoshida \& Lee 2000).
For $\bar\Omega=0.05$, the peak frequency $\bar\omega_P$ is halved compared to that for $\bar\Omega=0.1$.

\begin{figure}
\resizebox{0.33\columnwidth}{!}{
\includegraphics{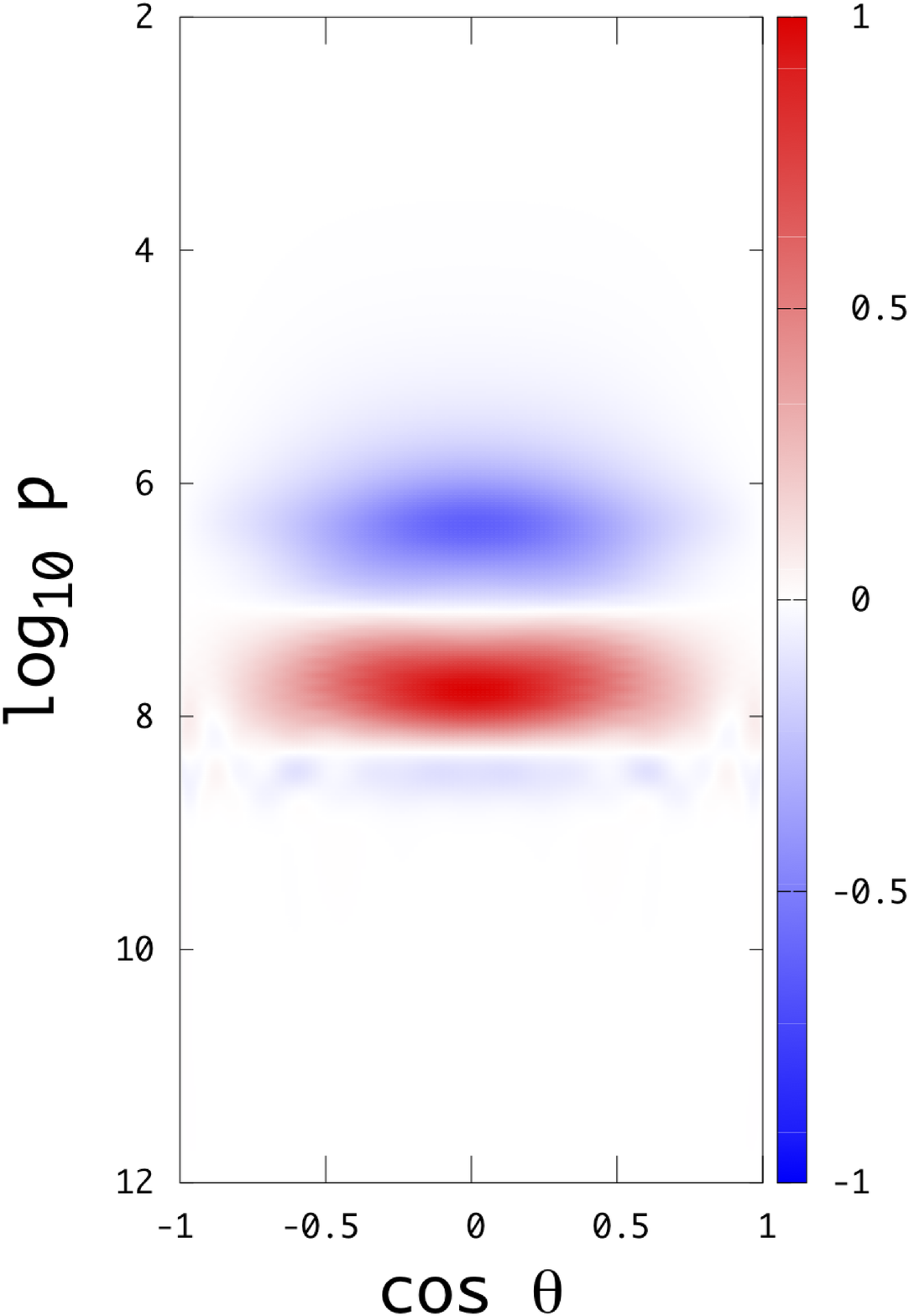}}
\resizebox{0.33\columnwidth}{!}{
\includegraphics{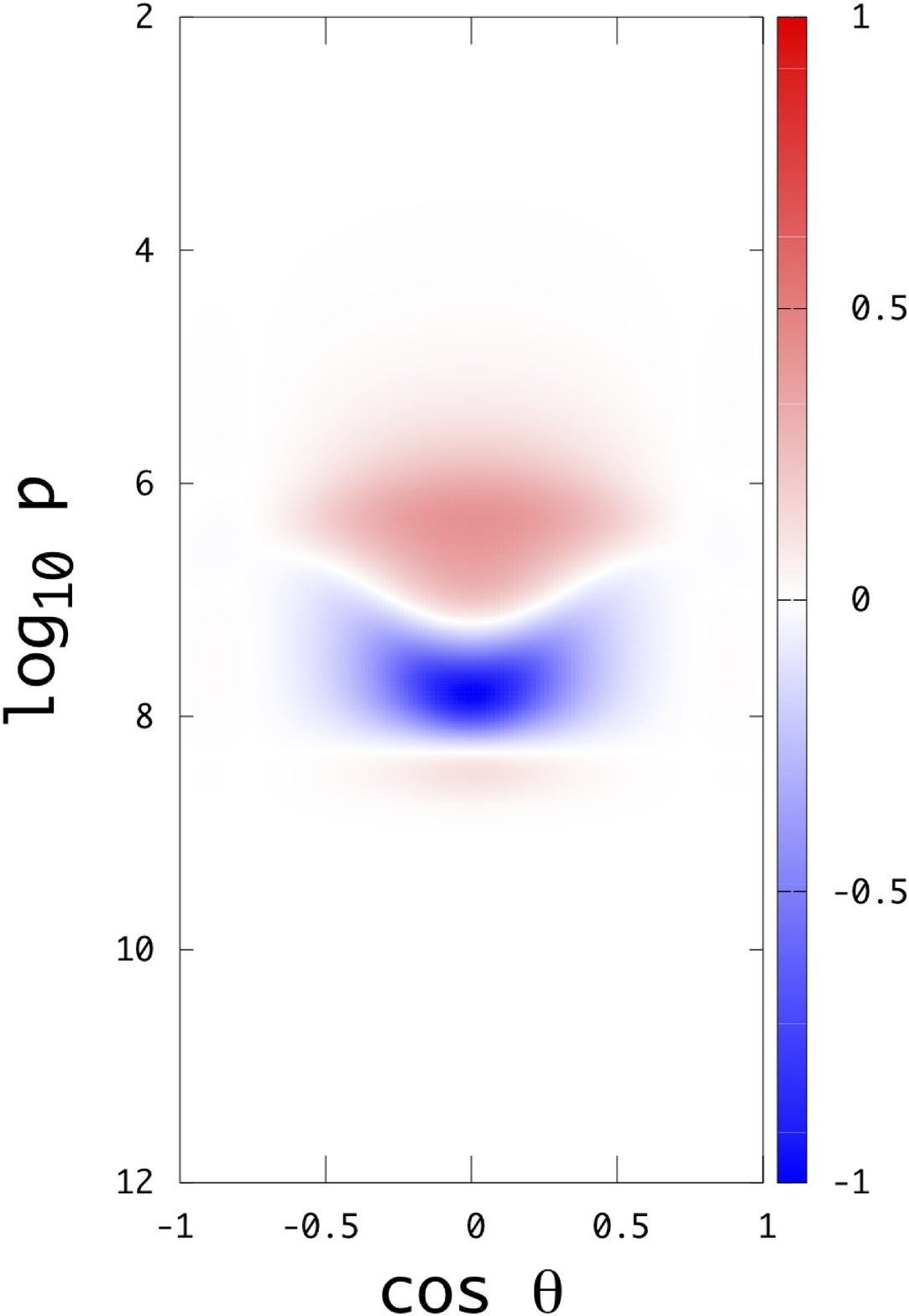}}
\resizebox{0.33\columnwidth}{!}{
\includegraphics{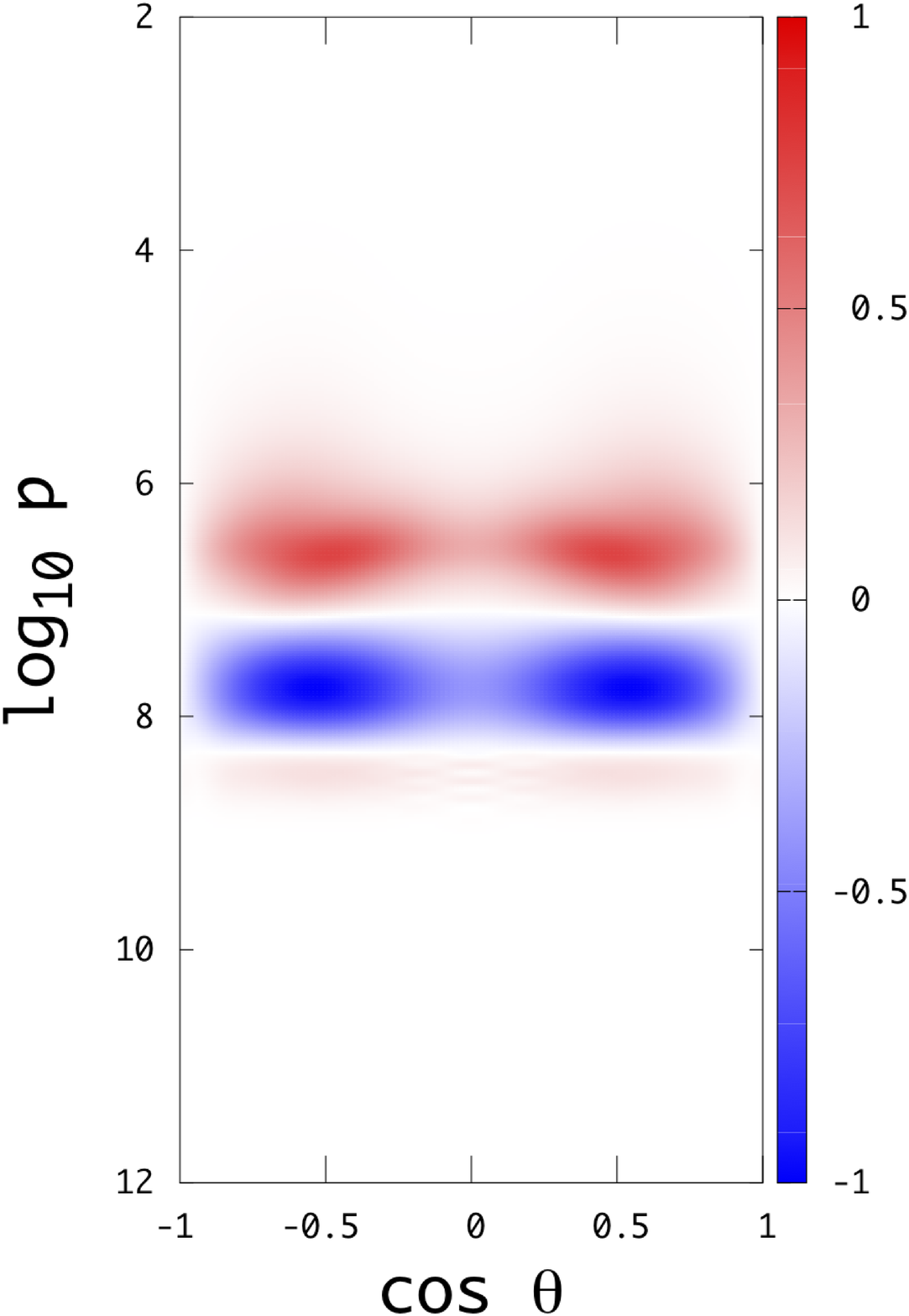}}
\caption{Color maps of $-{\rm Im}(\rho^\prime)$ for $\phi=0$
produced by semi-diurnal thermal tides, from left to right panels, at the forcing frequency tabulated in Table 2 for the prograde and retrograde $g_1$-modes and the $r_1$-mode for $\bar\Omega=0.1$, where the pressure $p$ is given 
in dyn/cm$^2$ and the magnitudes of the torque are normalized by the maximum value.
}
\end{figure}

\begin{figure}
\resizebox{0.33\columnwidth}{!}{
\includegraphics{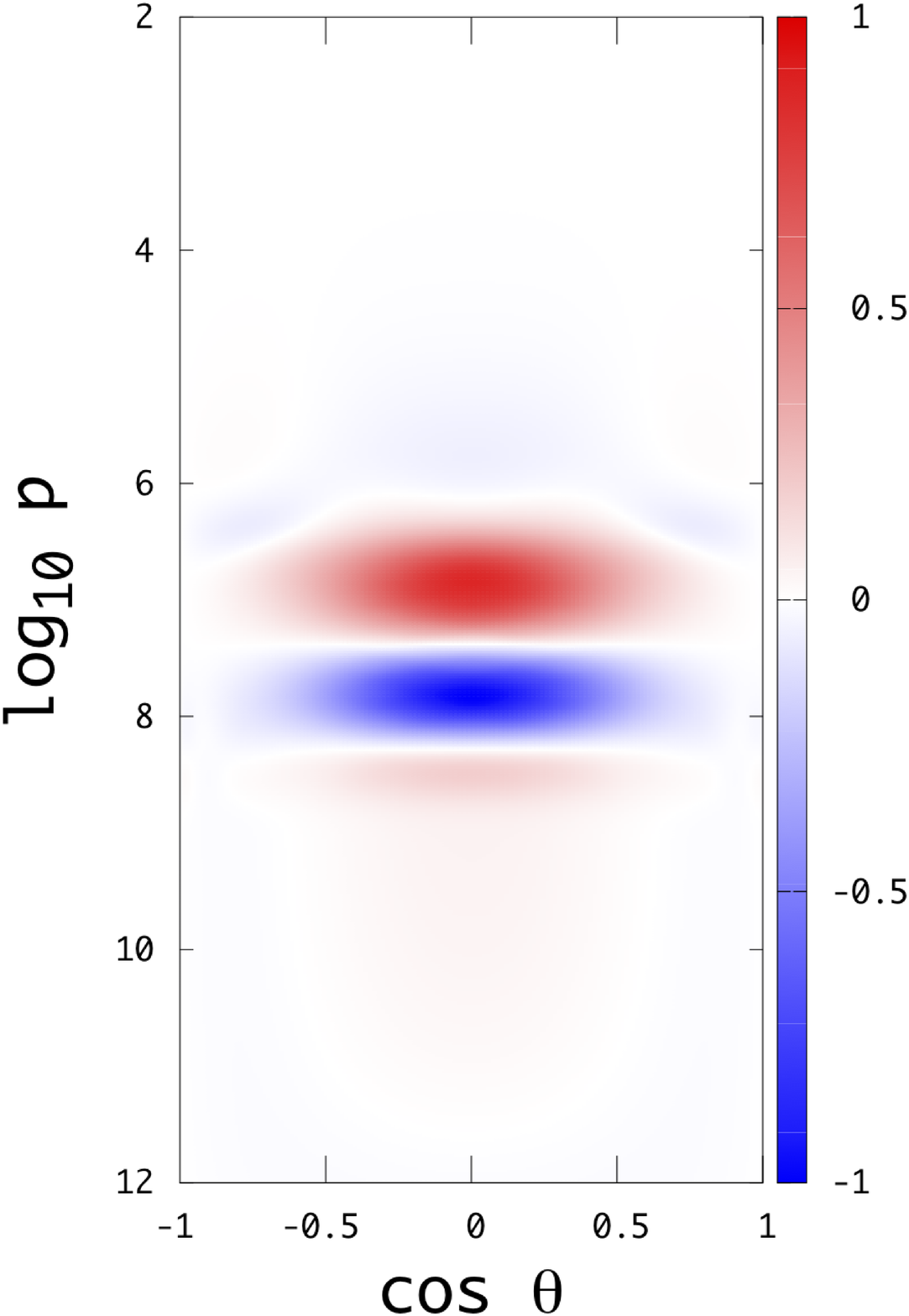}}
\resizebox{0.33\columnwidth}{!}{
\includegraphics{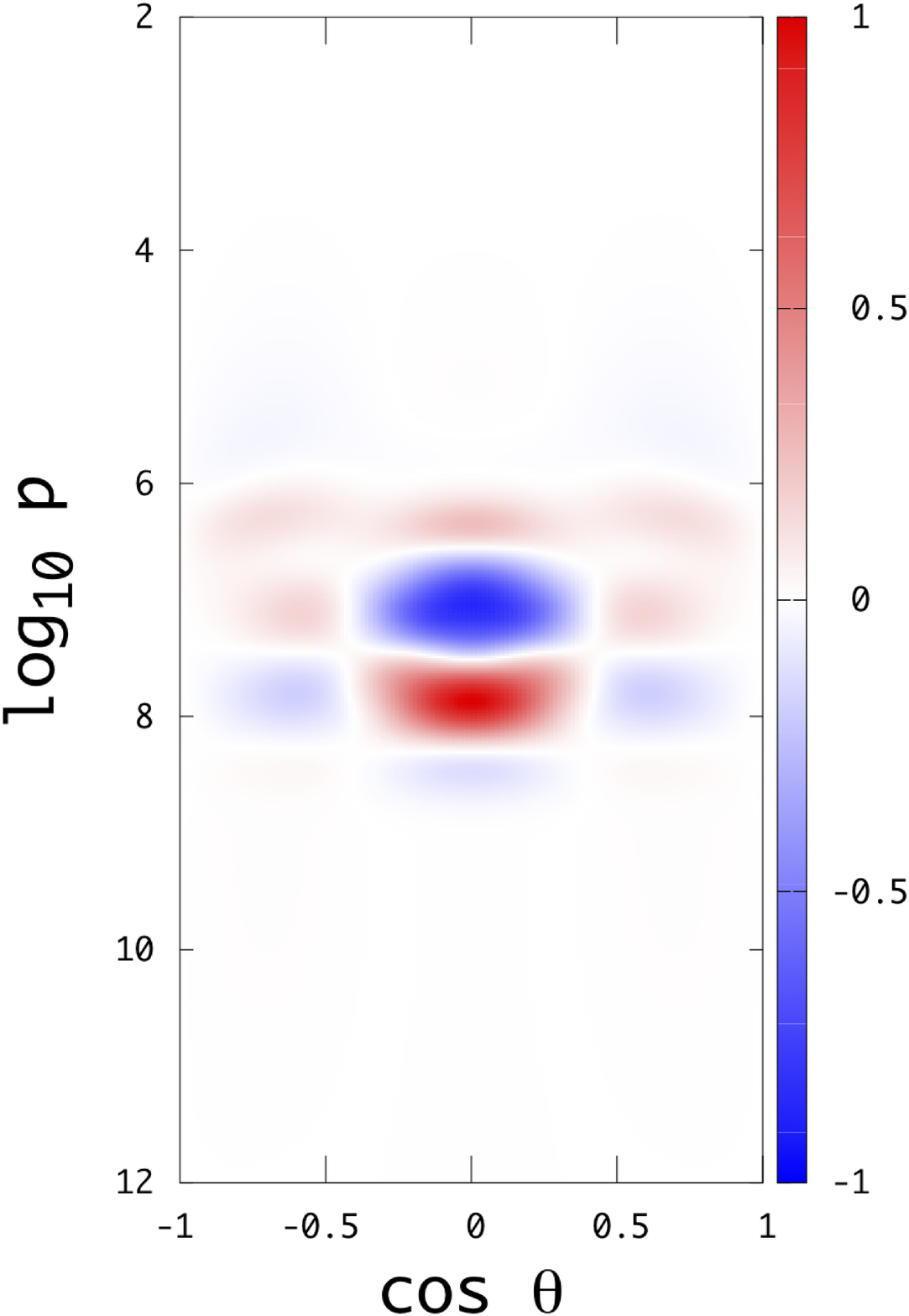}}
\caption{Same as Figure 5 but at the forcing frequency $\bar\omega=0.0557$ and $-0.11$, respectively
corresponding to the prograde and retrograde inertial modes $i_2$ in the core.
}
\end{figure}

The local tidal torque applied to a spherical surface is proportional 
to $-\Phi_T{\rm Im}(\rho^{\prime *})$, where $\Phi_T$ is given by equation (\ref{eq:phit}), and 
$\rho^\prime(r,\theta,\phi)=\sum_{j=1}^{j_{\rm max}}\rho^\prime_{l_j}(r)Y_{l_j}^m(\theta,\phi)$
for $m=-2$ where $j_{\rm max}=12$ is used in this paper.
In Figure 5 \& 6, we show the color-maps of $-{\rm Im}(\rho^\prime)$ in the $\cos\theta-\log_{10}p$ plane,
assuming $\phi=0$.
Figure 5 is for the prograde and retrograde $g_1$-modes and the $r_1$-mode at the
forcing frequency tabulated in Table 2 for $\bar\Omega=0.1$ and Figure 6 for prograde and retrograde inertial modes $i_2$ at the forcing frequency $\bar\omega=0.0557$ and $-0.11$, respectively.
The patterns are symmetric about the equator $\cos\theta=0$.
The local tidal torque is confined into a geometrically very narrow region at the bottom of the radiative envelope
and the direction of the torque changes in this narrow layer, which could lead to a strong differential
rotation there.
The amplitudes of the torque is confined in an equatorial region for the $g$-modes, and this confinement
is stronger for the retrograde $g_1$-mode.
At the forcing frequency of the $r_1$-mode, the amplitude has two peaks as a function of $\cos\theta$ and is small
at the equator.
At the resonant forcing frequency for the $i_2$ inertial modes in the core, the amplitude distribution 
for the retrograde inertial mode is much more complicated than that for the prograde inertial mode, which
has a similar distribution to that of the prograde $g_1$-mode.

Figure 7 shows the tidal torque ${\cal N}$ computed assuming $\pmb{j}_*=0$ and $\pmb{\psi}\not=0$
for $\tau_*=1$day. 
There appears more sharp peaks produced by resonance between the forcing and inertial modes in the core,
compared to the case of pure thermal tides.
The broad peaks due to the $g$-mode resonance are pierced by such sharp peaks due to the inertial modes.
Because the tidal potential $\Phi_T$ has substantial amplitudes in the convective core,
the inertial modes in the core are more susceptible to the gravitational tides than
the thermal tides.
We also find peaks due to the resonance with the envelope $r$-modes on the retrograde side.
%{\bf 
See the Appendix C for a discussion about the alternative changing of the sign of $\cal N$. %}

\begin{figure}
\resizebox{0.45\columnwidth}{!}{
\includegraphics{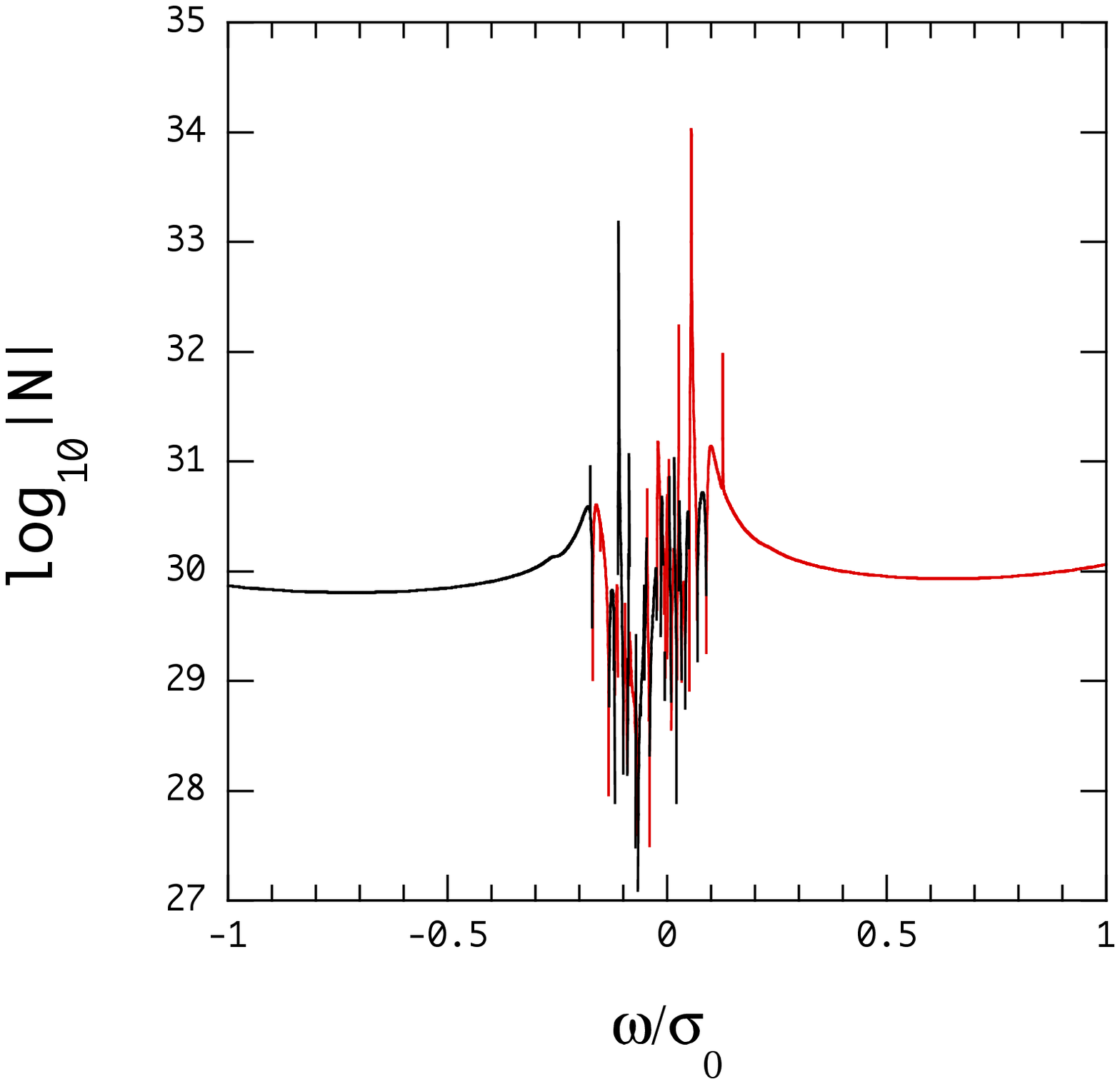}}
\hspace*{0.5cm}
\resizebox{0.45\columnwidth}{!}{
\includegraphics{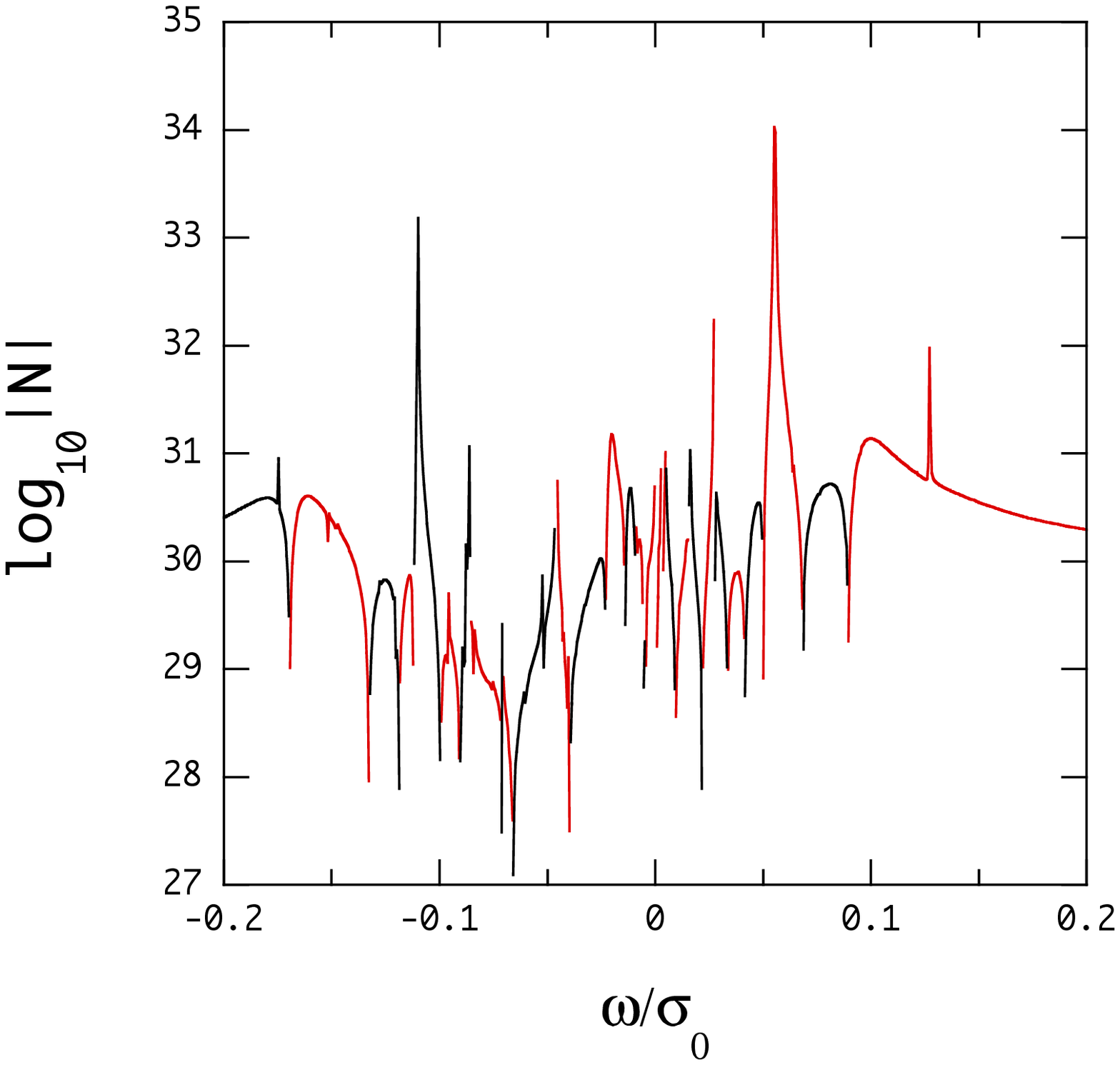}}
\caption{Same as Figure 4 but for $\pmb{j}_*=0$ and $\pmb{\psi}\not=0$.
}
\end{figure}

Instead of assuming the rotation rate $\bar\Omega$ takes a constant value,
we let $\bar\Omega$ change as a function of $\bar\omega$ (or $\bar\omega$ changes as a function of $\bar\Omega$) for a given $\bar\Omega_{\rm orb}$, that is,
$\bar\Omega$ is given by
\be
\bar\Omega=\bar\Omega_{\rm orb}-\bar\omega/2.
\ee
In Figure 8, we plot the tidal torque $\cal N$ as a function of the forcing period $\tau_{\rm tide}=2\pi/\omega$, where the left panel is for $\bar\Omega_{\rm orb}=0.0537$ and the right panel for 
$\bar\Omega_{\rm orb}=-0.0537$.
Since we assume $\tau_{\rm tide}>0$, $\bar\Omega$ changes sign for $\bar\Omega_{\rm orb}=0.0537$
but it stays negative for $\bar\Omega_{\rm orb}=-0.0537$.
Since prograde (retrograde) forcing corresponds to positive (negative) $\nu=2\Omega/\omega$,
as $\tau_{\rm tide}$ increases, the forcing changes from retrograde to prograde for
$\bar\Omega_{\rm orb}=0.0537$ and it is always retrograde for $\bar\Omega_{\rm orb}=-0.0537$.
We find the gross properties of $\cal N$ as a function of $\tau_{\rm tide}$
shown by the left panel of Figure 8 is similar to those computed by Auclaire-Desrotour \& Leconte (2018)
using the traditional approximation, except that we have sharp resonance peaks due to
inertial modes in the convective core.
The reasons for the difference may be partly because they used the traditional approximation, with
which inertial modes cannot be properly calculated, 
and partly because they assumed $\Gamma_1=1.4$, for which 
the convective core is not necessarily isentropic and propagation of inertial modes in the core may be suppressed.
Assuming negative $\bar\Omega_{\rm orb}$ (right panel), 
we can calculate retrograde forcing with long periods, with which the envelope $r$-modes are excited
for $\tau_{\rm tide}\gtsim 10$ days.

\begin{figure}
\resizebox{0.45\columnwidth}{!}{
\includegraphics{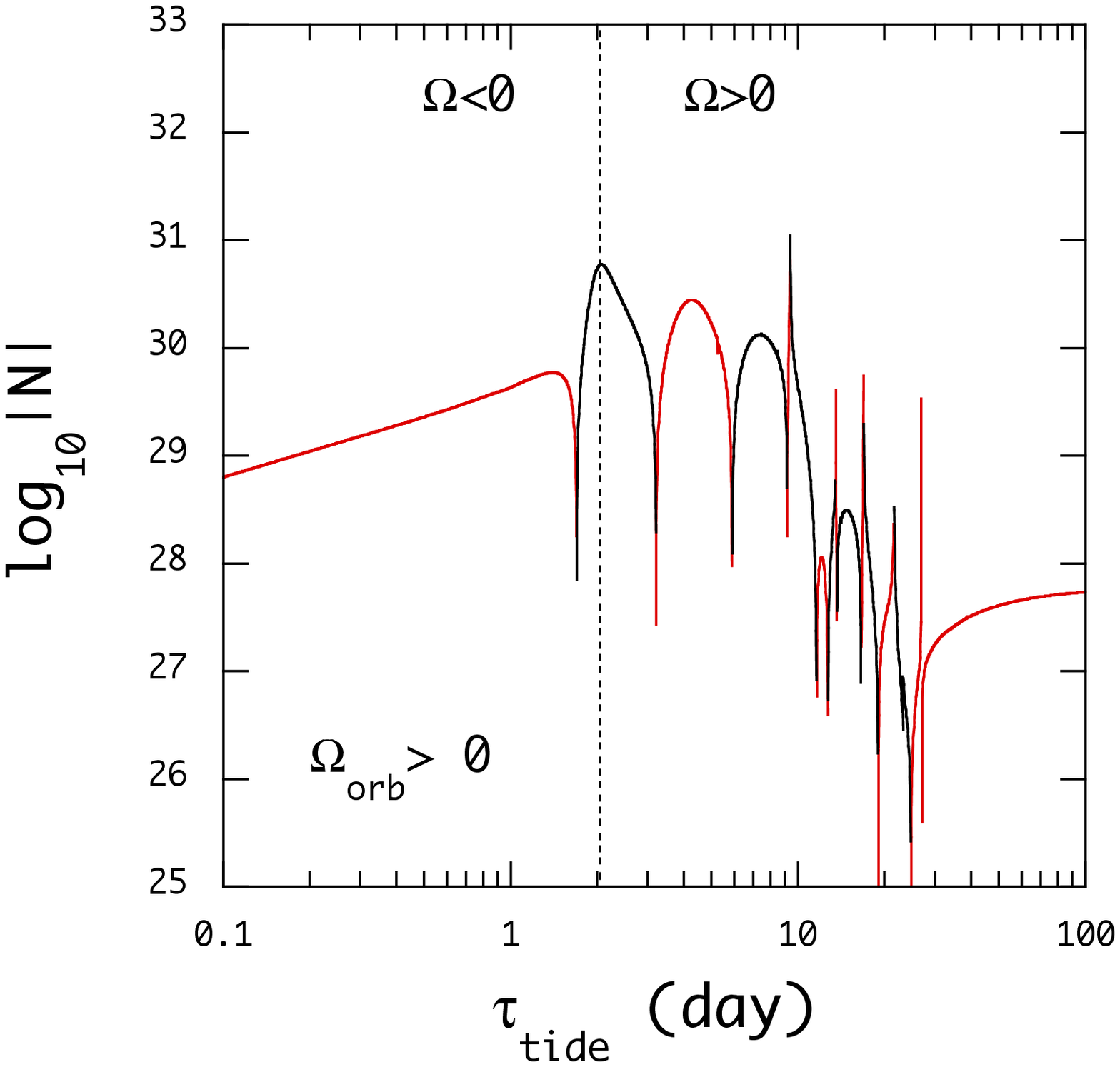}}
\hspace*{0.5cm}
\resizebox{0.45\columnwidth}{!}{
\includegraphics{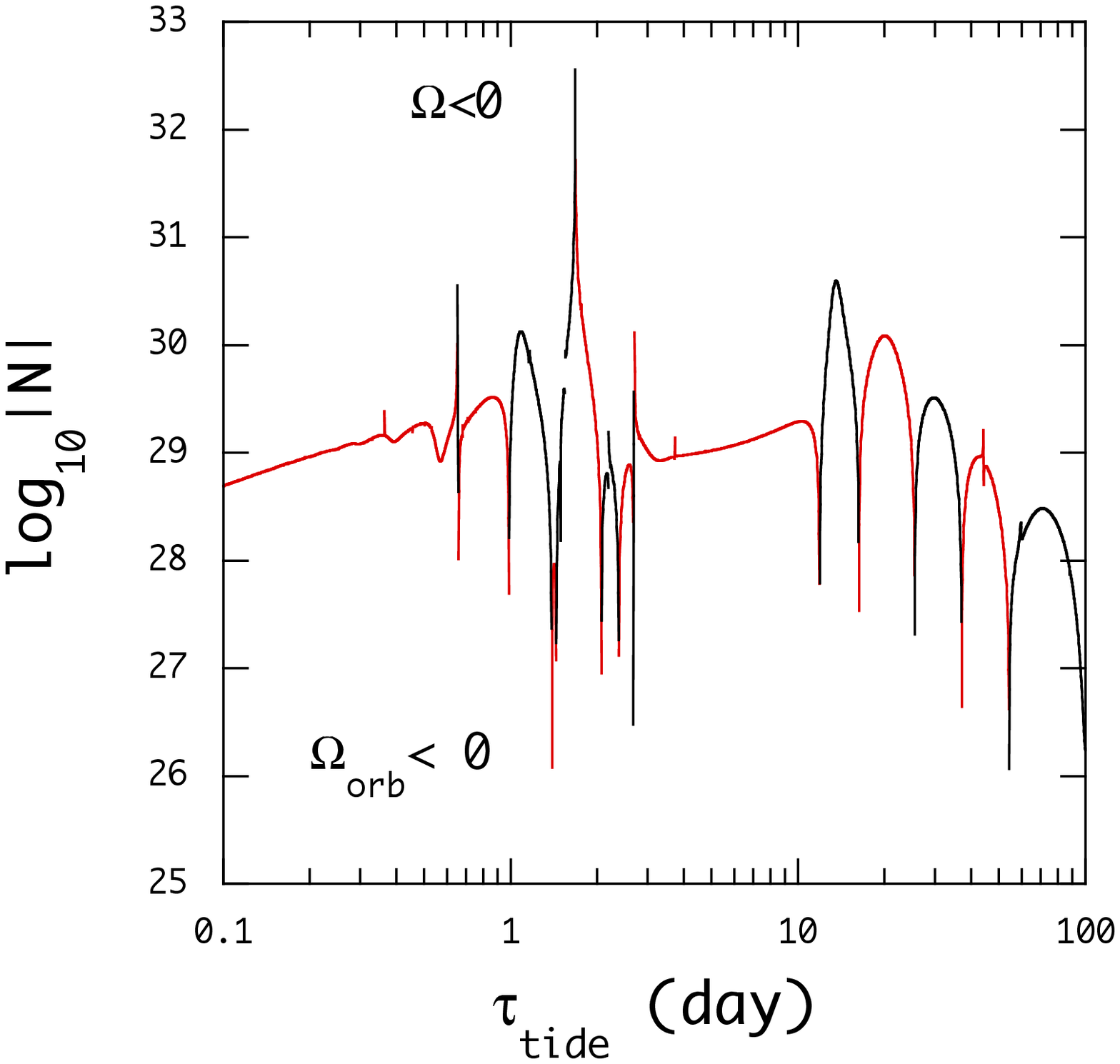}}
\caption{Tidal torque, given in erg, due to thermal tides for $\tau_*=1$ day
versus the tidal forcing period $\tau_{\rm tide}=2\pi/\omega$ in days,
where the rotation speed $\Omega$ of the planet is given by $\Omega=\Omega_{\rm orb}-\pi/\tau_{\rm tide}$ as a function of $\tau_{\rm tide}$ for a given $\Omega_{\rm orb}$, and
we use $\bar\Omega_{\rm orb}=0.0537$ for the left panel and $\bar\Omega_{\rm orb}=-0.0537$ for the right panel.
Here, the red lines and black lines respectively indicate positive and negative torque $\cal N$.
The vertical dotted line in the left panel indicates the forcing period at which $\bar\Omega=0$.
Note that $\bar\Omega<0$ ($\bar\Omega>0$) corresponds to the retrograde (prograde) forcing for $\tau_{\rm tide}>0$.
}
\label{fig:torque_period}
\end{figure}

\section{Conclusion}

We have computed the tidal torque due to thermal tides in rotating hot Jupiters, composed of a thin isothermal radiative envelope and a nearly isentropic convective core.
The thin envelope suffers the strong irradiation by the host star
and the periodic alternations of day and night sides on the planet produces semi-diurnal thermal tides.
We have taken into consideration radiative cooling in the envelope as the non-adiabatic energy dissipation mechanism.
To represent the tidal responses in rotating planets, we use series expansions in terms of 
spherical harmonic functions $Y_l^m(\theta,\phi)$ with different $l$s for a given $m$.
For fixed values of $\bar\Omega$, we have computed the tidal torque as a function of the tidal forcing frequency $\omega$ for both prograde and retrograde forcing, observed in the co-rotating frame of the planet.
We find that at the forcing frequency $|\omega|\sim\sqrt{GM/R^3}$, the tidal torque tends to
synchronize the planet spin with the orbital motion, the direction of which is the same as
that by gravitational tides.
At low frequency, the tidal forcing can be in resonance with low frequency modes
such as $g$-modes and $r$-modes in the envelope and inertial modes in the core
and the resonance tends to enhance the tidal torques.
The sign of the tidal torque at the resonance peaks changes alternately as the mode that is in resonance
with the forcing is changed with $|\omega|$.
The tidal resonance with the $g$- and $r$-modes produces broad peaks of the torque and that with
the inertial modes sharp peaks as a function of the forcing frequency.
The peak frequency $\omega_P$ of the broad peaks by the $g$- and $r$-modes is only weakly dependent on the spin frequency $\Omega$ and $\omega_P$ of 
the sharp peaks is proportional to $\Omega$.

We find a few differences between the results obtained in this paper and those
by Auclair-Desrotour \& Leconte (2018).
One of the differences may concern the core inertial modes of rotating Jovian planets.
The traditional approximation employed by Auclair-Desrotour \& Leconte (2018) 
to represent the tidal responses in rotating planets cannot properly treat inertial modes
propagating isentropic regions.
The Jovian models used in this paper and by Auclair-Desrotour \& Leconte (2018) have the convective core
that has the structure of a polytrope of the index $n=1$.
For the core Auclair-Desrotour \& Leconte (2018) assumed $\Gamma_1=1.4$ to avoid a nearly
isentropic structure and hence suppressed core inertial modes.
On the other hand, we use series expansion in terms of spherical harmonic functions to represents
tidal responses in rotating planets and assume $\Gamma_1=2$ to make the core nearly isentropic, which
supports propagation of inertial modes.
Because core inertial modes are not necessarily susceptible to thermal tides 
prevailing in the radiative envelope, the difference between the present study and 
Auclair-Desrotour \& Leconte (2018) concerning the inertial modes
may be considered as a minor difference, although at the resonance peak with inertial modes
the magnitude of tidal torque is significantly enhanced.
Another difference between the two analyses may concern the resonance with the envelope $r$-modes.
Assuming $\Omega_{\rm orb}<0$, in this paper, we could compute retrograde forcing with long periods
and hence the tidal torques in resonance with the envelope $r$-modes.
As shown by Figure 8, the behavior of tidal torques as a function of the forcing period $\tau_{\rm tide}$
is different between prograde and retrograde forcing with long periods, that is,
as $\tau_{\rm tide}$ increases the tidal torque on the
prograde side stay positive to work for synchronization but on the retrograde side
it changes its sign alternatively.

\begin{figure}
\resizebox{0.45\columnwidth}{!}{
\includegraphics{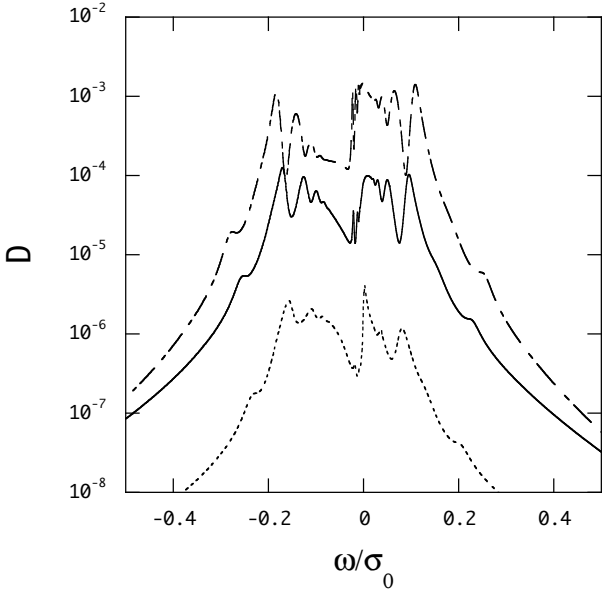}}
\caption{Normalized energy dissipation rate $D$ due to thermal tides versus forcing frequency $\bar\omega$
for $\tau_*=10$day (dash-dotted line), 1day (solid line), and 0.1day (dotted line) for
$\bar\Omega=0.1$ where $D$ is defined by equation (\ref{eq:edr}).
}
\end{figure}

We compute the rate of energy dissipation cause by thermal tides in the envelope where non-adiabatic effects
are significant.
We define the normalized energy dissipation rate $D$ as
\be
D
={\bar\omega\over 2}\int_0^R{\rm Im}\left(\sum_l
{\delta T^*_l\over T}{\delta s_l\over c_p}\right){\rho T c_pr^3\sigma_0\over L_{eq}}{dr\over r},
\label{eq:edr}
\ee
where $L_{eq}\equiv 4\pi R^2F_*=4\pi R^2\sigma_{\rm SB}T_*^4(R_*/r_*)^2$ (see Lee 2019).
For the stellar parameters we use in this paper, we have $L_{eq}\approx 6\times10^{29}{\rm erg/s}$
and hence $L_{extra}/L_{eq}\sim10^{-2}$ for $L_{extra}\sim 10^{27}-5\times10^{27}{\rm erg/s}$, which
is the magnitude of the extra heat source needed to inflate the planets
(see Baraffe et al 2003).
As suggested by Figures 5 \& 6, strong heating due to thermal tides occurs in the bottom layers of the envelope.
In Figure 9, we plot $D$ as a function of the forcing frequency $\bar\omega$ for three values of $\tau_*$ for $\bar\Omega=0.1$.
The dissipation rate has large values for $|\bar\omega|\ltsim 0.2$, corresponding to 
the frequency range in which tidal forcing can be in resonance with the low frequency modes in the envelope and inertial modes in the core.
The magnitude of $D$ increases as $\tau_*$ increases and it becomes $D\sim 10^{-3}$
for $\tau_*=10$day, suggesting that non-adiabatic heating caused by thermal tides 
at the bottom of the envelope can be a heating source for inflation of the planets if
$\tau_*$ is sufficiently long.

%{\bf 
The results presented in this paper may depend on the expansion length $j_{\rm max}$ if the length is not long enough.
We compute for $j_{\rm max}=20$ the tidal torque as a function of the forcing frequency assuming $\pmb{j}_*=0$ and $\pmb{\psi}\not=0$, and
the result is shown by Figure 10.
Comparing to Fig. 7, for which we assumed $j_{\rm max}=12$, we find that
the tidal torque as a function of the forcing frequency $\bar\omega$ is almost the same between the cases of 
$j_{\rm max}=12$ and 20.
We confirm that the length $j_{\rm max}=12$ is long enough to produce reliable results. %}

\begin{figure}
\resizebox{0.45\columnwidth}{!}{
\includegraphics{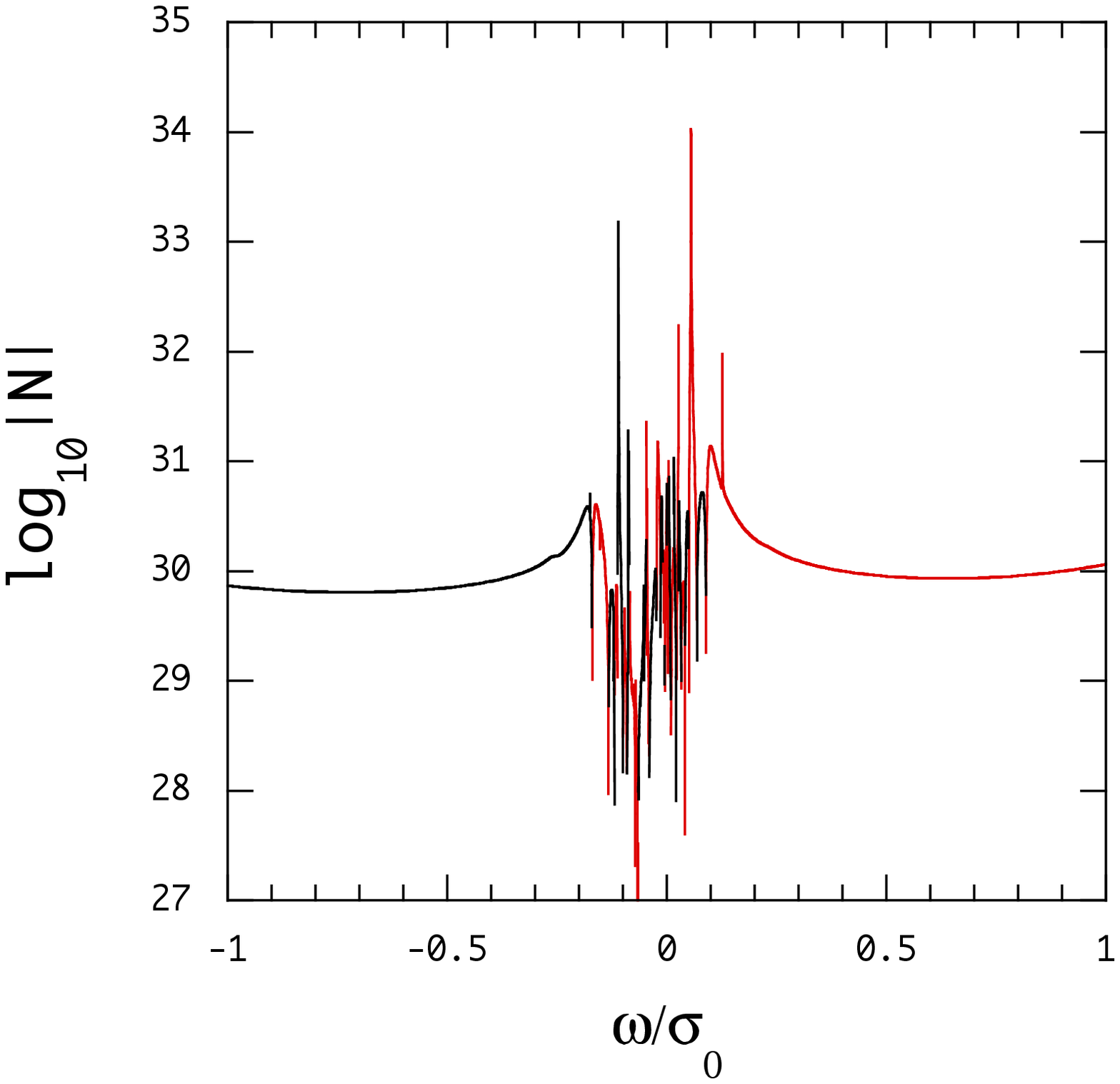}}
\hspace*{0.5cm}
\resizebox{0.45\columnwidth}{!}{
\includegraphics{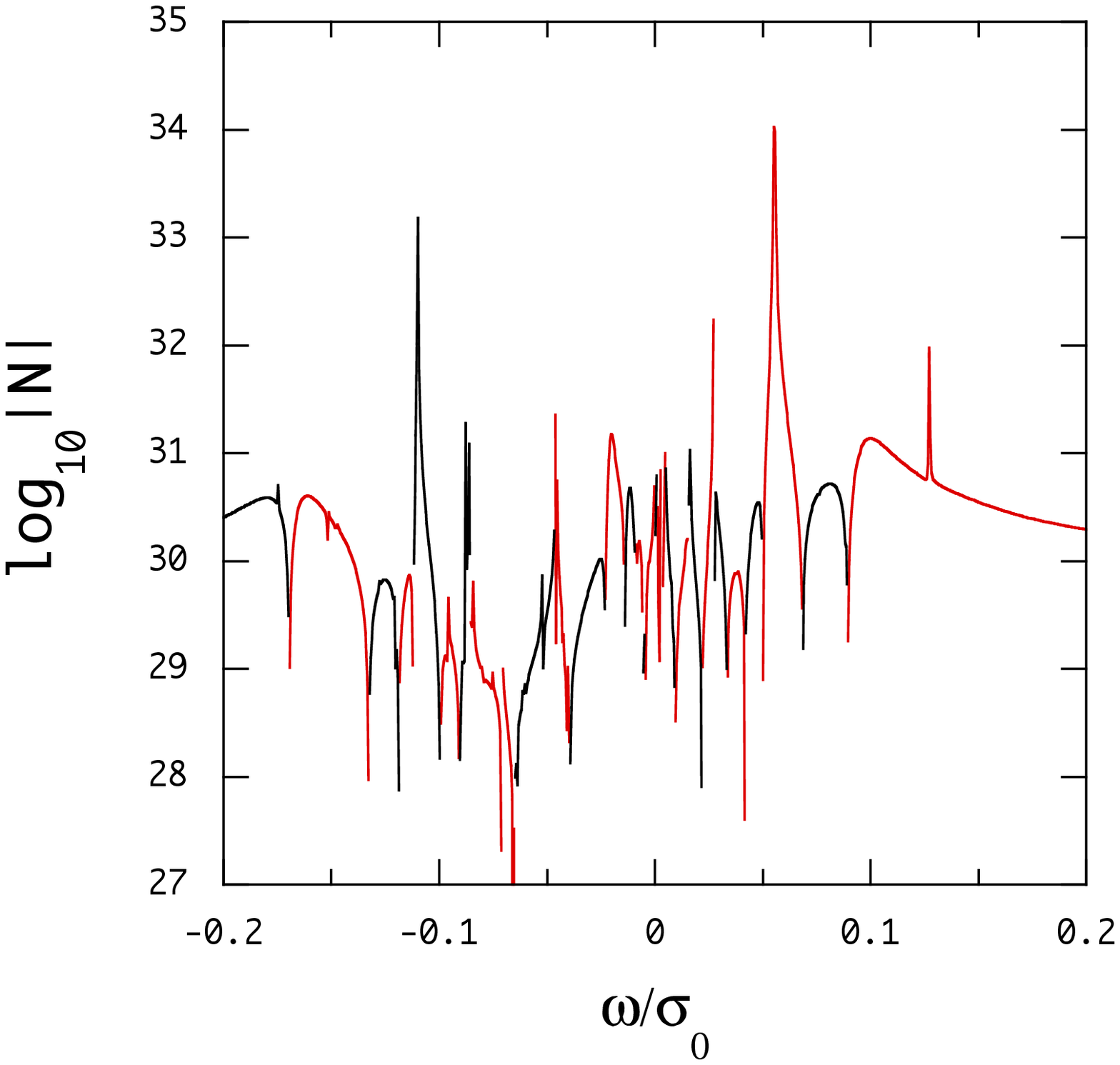}}
\caption{Same as Figure 7 but for $j_{\rm max}=20$.
}
\end{figure}

\begin{figure}
\resizebox{0.45\columnwidth}{!}{
\includegraphics{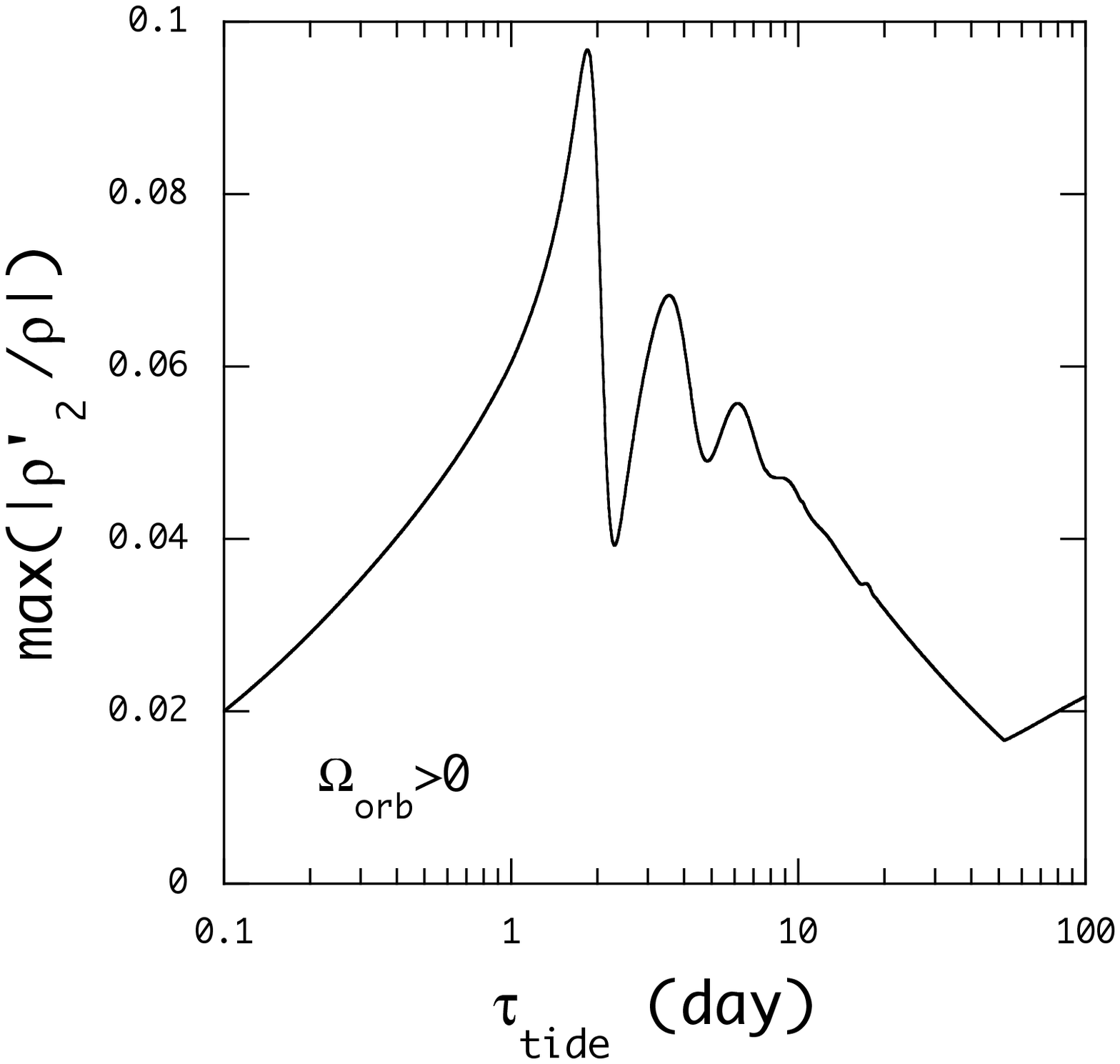}}
\hspace*{0.5cm}
\resizebox{0.45\columnwidth}{!}{
\includegraphics{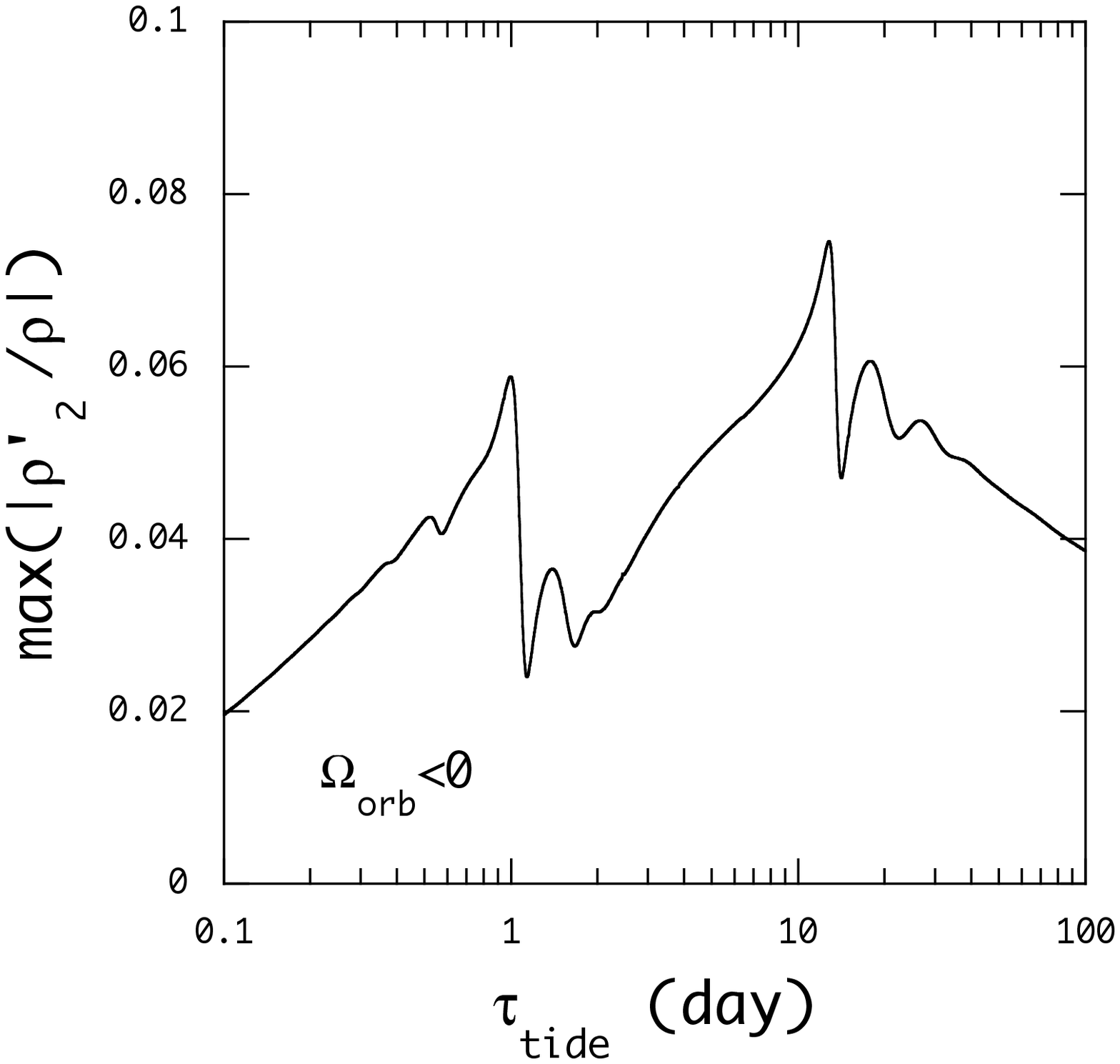}}
\caption{${\rm max}\left(|\rho'_2/\rho|\right)$ as a function of the forcing period $\tau_{\rm tide}$ (day) for 
$\pmb{j}_*\not=0$ and $\pmb{\psi}=0$ and for $\tau_*=1$day, where
the rotation speed $\Omega$ is given by $\Omega=\Omega_{\rm orb}-\pi/\tau_{\rm tide}$.
This figure corresponds to Fig. \ref{fig:torque_period}.
}
\label{fig:rhomax}
\end{figure}

%{\bf 
With an asymptotic treatment of waves, the local strength of nonlinearity of the waves
could be discussed by using a quantity $\xi_r k_r$, where $k_r$ is the radial component of the wavenumber vector
(e.g., Goodman \& Dickson 1998).
Here instead we simply use the quantity ${\rm max}\left(|\rho'_2/\rho|\right)$, which is the maximum value of $|\rho'_2/\rho|$
in the interior of the planet, to consider the validity of linear approximation employed in this paper.
Here $-{\rm Im}(\rho'_2)$ is used to compute the tidal torque.
In Fig. \ref{fig:rhomax} we plot ${\rm max}\left(|\rho'_2/\rho|\right)$ as a function of the forcing period $\tau_{\rm tide}$ (day) 
for $\pmb{j}_*\not=0$ and $\pmb{\psi}=0$, which corresponds to Fig. \ref{fig:torque_period}, where for a given value of $\bar\Omega_{\rm orb}$
the rotation speed $\Omega$ is given by $\Omega=\Omega_{\rm orb}-\pi/\tau_{\rm tide}$ as a function of $\tau_{\rm tide}$.
This figure shows that the amplitudes of the tidal responses to pure thermal tides are less than 0.1 and stay in a linear regime
for the parameters used in this paper.
So long as the amplitudes stay in a linear regime, the amplitudes are proportional to the external parameter $F_*$, which depends on
the luminosity of the host star and the distance between the host star and the planet.
As suggested by the figure, however, if the parameter $F_*$ increases by one or two order of magnitudes, the responses to the thermal tides
enter into a non-linear regime and we need non-linear treatment of the responses.
In a nonlinear regime, the tidal responses excite many different oscillation modes by non-linear mode coupling,
leading to a strong damping of the responses (e.g., Kumar \& Goodman 1996).
Note that for pure gravitational tides ($\pmb{j}_*=0$ and $\pmb{\psi}\not=0$), we already have
${\rm max}\left(|\rho'_2/\rho|\right)\sim 1$, suggesting that we need non-linear treatment of the responses, 
although the amplitudes are quite uncertain because we consider no dissipative processes in the convective core in which
the perturbing tidal potential $\Phi_T$ has substantial amplitudes. %}

It is useful to make clear the relation between the methods of solutions used in this paper and by
Auclair-Desrotour \& Leconte (2018) for thermal tides.
To represent the tidal responses in rotating planets, we use series expansions in terms of
spherical harmonic functions $Y_l^m(\theta,\phi)$.
The tidal torque on the planet, if we simply assume the tidal potential given by 
$\Phi_T\propto Y_2^{-2}$,
may be given by equation (\ref{eq:tidaltorque})
and $\rho'^*_2$ in this equation is obtained by solving equations (\ref{eq:dy1}) and (\ref{eq:dy2}) and using equation (\ref{eq:cont}).
Auclair-Desrotour \& Leconte (2018), on the other hand,
used series expansions of the responses in terms of the Hough functions $\Theta_{km}(\theta;\nu)$ defined in the traditional
approximation (e.g., Lee \& Saio 1997).
Defining $\hat\Theta_{km}(\theta,\phi;\nu)=f_{km}\Theta_{km}e^{\rmi m\phi}$ where $f_{km}$
is introduced so that the normalization $\left<\hat\Theta_{k'm}|\hat\Theta_{km}\right>=\delta_{k'k}$
is satisfied and
\be
\left<f| g\right>=\int_0^\pi d\theta\sin\theta\int_0^{2\pi}d\phi f^*g,
\ee
we may have, assuming the functions $\hat\Theta_{km}$ form a complete set,
\be
Y_l^m=\sum_ky_{lk}\hat\Theta_{km}, \quad \rho'(r,\theta,\phi)=\sum_k\hat\rho'_k(r)\hat\Theta_{km}.
\ee
For the density perturbation $\rho^\prime$ given by $\rho'_2Y_2^{-2}$, we obtain
\be
\rho'_2=\left<Y_2^{-2}|\rho^\prime\right>=\sum_k\hat\rho'_k\left<Y_2^{-2}|\hat\Theta_{k,-2}\right>=\sum_k\hat\rho'_ky_{2k}^*,
\ee
and for the tidal potential $\Phi_T=\Phi_T(r) Y_2^{-2}=\sum_k\hat\Phi_{T,k}\hat\Theta_{k,-2}$
\be
\hat\Phi_{T,k}=\left<\hat\Theta_{k,-2}|\Phi_T\right>=\Phi_T(r)y_{2k}.
\ee
Adding $\hat\Phi_{T,k}$ as an inhomogeneous forcing term, we compute
the density perturbations $\hat\rho'_k$ in the traditional approximation.

The tidal torque due to equilibrium gravitational tides may be estimated as (e.g., Goldreich \& Soter 1966)
\be
{\cal N}_{\rm eq}={3GM_*^2R^5\over 2a_*^6}{1\over Q},
\ee
where $Q$ is the tidal quality factor, representing the magnitude of 
the phase lag caused by energy dissipations
that arise from interaction between the tidal potential and fluid motion in the interior.
The $Q$ value for the interaction between the tidal potential and the convective core 
is difficult to estimate since the fluid motion in the core is usually turbulent
so that we need properly treat effective viscosity for turbulence 
to estimate the amount of energy dissipations
(see, e.g., Zahn 1977; Goldreich \& Nicholson 1977).
For the parameters used in this paper, we have ${\cal N}_{\rm eq}=1.6\times10^{38}/Q$, 
which could be comparable to the torque due to the thermal tides calculated in this paper
only for $Q\gtsim 10^7$, except for those at the peaks produced by resonance with
inertial modes.
Probably, the magnitude $Q\gtsim 10^7$ is too large for Jovian planets (e.g., Goldreich \& Nicholson 1977).
As discussed by Auclair-Desrotour \& Leconte (2018), if we consider local timescales for the rotation rates to
change in the envelope and in the convective core,
the two timescales can be comparable with each other for reasonable values of $Q$
since the moment of inertia of the thin envelope is much smaller that that of the convective core.
If we assume certain formulae for turbulent viscosity coefficient as done by
Ogilvie \& Lin (2004), we could estimate the tidal torque caused by both gravitational
and thermal perturbations although we have to solve the Navier Stokes equations for
rotating planets, which will be one of our future works.

\begin{appendix}

\section{Derivation of the Oscillation Equations}

In this Appendix, we give a brief account of the derivation of the oscillation equations (30) to (34).
The three components of the perturbed equation of motion (\ref{eq:eqmot}) are written as
\be
-\rho\omega^2\xi_r-2\rmi \omega\Omega\rho\xi_\phi\sin\theta=-{\partial p^\prime\over\partial r}
-\rho'{d\Phi\over dr}-\rho^\prime{\partial\Phi_T\over\partial r},
\label{eq:eom_r}
\ee
\be
-\omega^2\rho\xi_\theta-2\rmi \omega\Omega\xi_\phi\cos\theta=-{1\over r}{\partial p'\over \partial\theta}-\rho{1\over r}{\partial \Phi_T\over \partial\theta},
\label{eq:eom_theta}
\ee
\be
-\omega^2\rho\xi_\phi+2\rmi\omega\Omega\rho\left(\xi_\theta\cos\theta+\xi_r\sin\theta\right)=-{1\over r\sin\theta}{\partial p'\over\partial\phi}-\rho{1\over r\sin\theta}{\partial \Phi_T\over\partial\phi}.
\label{eq:eom_phi}
\ee
Substituting the expansions given by (\ref{eq:prhoperturbations}) to (\ref{eq:disp_phi})
into equation (\ref{eq:eom_r}), we find that the radial component of the equation of motion (\ref{eq:eom_r}) reduces to
\begin{eqnarray}
\sum_l\left(-c_1\bar\omega^2S_l+2 c_1\bar\omega\bar\Omega mH_l\right)Y_l^m
+2c_1\bar\omega\bar\Omega\sum_{l'}\rmi T_{l'}\sin\theta{\partial\over\partial\theta} Y_{l'}^m
= \sum_l{P_l\over \rho g}Y_l^m,
\label{eq:eom2_r}
\end{eqnarray}
where 
\be
P_l=-\rho gr{\partial \over\partial r}Y_{2,l}
-\rho g{d\ln\rho gr\over d\ln r}Y_{2,l}
+\rho g{\Phi_{T,l}\over gr}{d\ln \rho\over d\ln r}-\rho g{\rho'_l\over\rho},
\ee
\be
Y_{2,l}={p'_l\over\rho gr}+{\Phi_{T,l}\over gr}.
\ee
Similarly, using the $\theta$ and $\phi$ components of the perturbed equation of motion, 
$\sin^{-1}\theta\partial_\theta\sin\theta({\rm eq.}~\ref{eq:eom_theta})
+\sin^{-1}\theta\partial_\phi({\rm eq.}~\ref{eq:eom_phi})$, which is the divergence of the horizontal displacement where
$\partial_\theta={\partial/\partial\theta}$ and $\partial_\phi={\partial/\partial\phi}$,
gives
\begin{eqnarray}
\sum_l\left(c_1\bar\omega^2\Lambda_lH_l-2 c_1\bar\omega\bar\Omega mH_l-2mc_1\bar\omega\bar\Omega S_l\right)Y_l^m 
- \sum_{l'}2 c_1\bar\omega\bar\Omega\left(\Lambda_{l'}\rmi T_{l'}\cos\theta+\rmi T_{l'}\sin\theta{\partial\over\partial\theta} \right)Y_{l'}^m
=\sum_l\Lambda_l{Y_{2,l}}Y_l^m,
\label{eq:div}
\end{eqnarray}
and 
$\sin^{-1}\theta\partial_\theta\sin\theta({\rm eq.}~\ref{eq:eom_phi})-\sin^{-1}\theta\partial_\phi({\rm eq.}~\ref{eq:eom_theta})$, which corresponds to the radial component of 
$\nabla\times\pmb{\xi}$, gives
\begin{eqnarray}
\sum_{l'}\left(-c_1\bar\omega^2\Lambda_{l'}\rmi T_{l'}+2 c_1\bar\omega\bar\Omega m\rmi T_{l'} \right)Y_{l'}^m
+2 c_1\bar\omega\bar\Omega\sum_l\left( \Lambda_lH_l\cos\theta +H_l\sin\theta{\partial\over\partial\theta}
 -2S_l\cos\theta -S_l\sin\theta{\partial\over\partial\theta}\right)Y_l^m=0.
\label{eq:rot}
\end{eqnarray}
The linearized continuity equation (\ref{eq:cont}) may reduce to
\be
\sum_l\left(\rho'_l+{1\over r^2}{\partial\over\partial r}r^3\rho S_l-\rho \Lambda_lH_l\right)Y_l^m=0,
\label{eq:cont_l}
\ee
and the entropy perturbation (\ref{eq:dsocp}) to
\be
\sum_l\left[{\delta s_l\over c_p}-{1\over\rmi\omega+\omega_D}{\epsilon'_l\over Tc_p}+{\omega_D\over\rmi\omega+\omega_D}\left(\nabla_{\rm ad}
{\delta p_l\over p}+\nabla VS_l\right)\right]Y_l^m=0.
\label{eq:dsocp_l}
\ee
Using the relations given by
\be
\sin\theta{\partial Y_l^m\over\partial\theta}=lJ_{l+1}^mY_{l+1}^m-(l+1)J_l^mY_{l-1}^m,
\ee
\be
\cos\theta Y_l^m=J_{l+1}^mY_{l+1}^m+J_l^mY_{l-1}^m,
\ee
where
$
J_l^m=\sqrt{(l^2-m^2)/ (4l^2-1)}
$
for $l\ge|m|$ and $J_l^m=0$ otherwise, we rewrite each of the equations (\ref{eq:eom2_r}), (\ref{eq:div}),
(\ref{eq:rot}), (\ref{eq:cont_l}), and (\ref{eq:dsocp_l}) into the form $\sum_lA_lY_l^m=0$.
With the dependent variables as defined by equation (\ref{eq:depvar}), each set of
the equations $A_{l_j}=0$ for $j=1,~\cdots, ~j_{\rm max}$ is written in the form as given by the oscillation equations (30) to (34).
Note that equations (\ref{eq:eom2_r}), (\ref{eq:div}),
(\ref{eq:rot}), (\ref{eq:cont_l}), and (\ref{eq:dsocp_l}) correspond to equations (30), (33),
(32), (31), and (34), respectively.

\section{Inner boundary conditions}

At the centre of the planet, the set of linear ordinary differential equations (\ref{eq:dy1}) and (\ref{eq:dy2}) can be formally
written as
\be
r{d\pmb{z}\over dr}=\pmbmt{A}\pmb{z}, \quad \pmb{z}=(z_j)=\left(\begin{array}{c} \pmb{y}_1\\ \pmb{Y}_2\end{array}\right),
\label{eq:b1}
\ee
where $\pmbmt{A}$ is the coefficient matrix for the differential equations and $z_j$ is for
$j=1,~\cdots,~2j_{\rm max}$
for the expansion length $j_{\rm max}$.
Assuming $\pmb{z}\propto r^\beta$ at the center and substituting into (\ref{eq:b1}) (see, e.g., Unno et al 1989), we obtain
\be
\left(\pmbmt{A}-\beta\pmbmt{I}\right)\pmb{z}=0,
\ee
which gives $2j_{\rm max}$ eigenvalues $\beta_j$ and eigenfunctions $\pmb{z}_j$.
Among the $2j_{\rm max}$ eigenvalues, we pick up $j_{\rm max}$ eigenvalues $\beta_j$ that satisfy 
the regularity condition given by ${\rm Re}(\beta_j)\ge-1$ and the corresponding eigenfunctions $\pmb{z}_j$.
Using these eigenvalues and eigenfunctions, we may represent the function $\pmb{z}$ at the centre as
\be
\pmb{z}=\sum_{j=1}^{j_{\rm max}}C_jr^{\beta_j}\pmb{z}_j=\pmbmt{Z}\left(\begin{array}{c}C_1r^{\beta_1}\\ \vdots\\
C_{j_{\rm max}}r^{\beta_{j_{\rm max}}}\end{array}\right), \quad
\pmbmt{Z}=(\pmb{z}_1, \cdots, \pmb{z}_{j_{\rm max}}),
\ee
where $C_j$ are arbitrary constants.
Eliminating the terms $C_jr^{\beta_j}$, we obtain $j_{\rm max}$ linear relations between $z_j$, which we use as the 
inner boundary conditions.

\section{Tidal torque $\cal N$ as a function of $\omega$ for $\pmb{\psi}\not=0$}

\begin{figure}
\resizebox{0.45\columnwidth}{!}{
\includegraphics{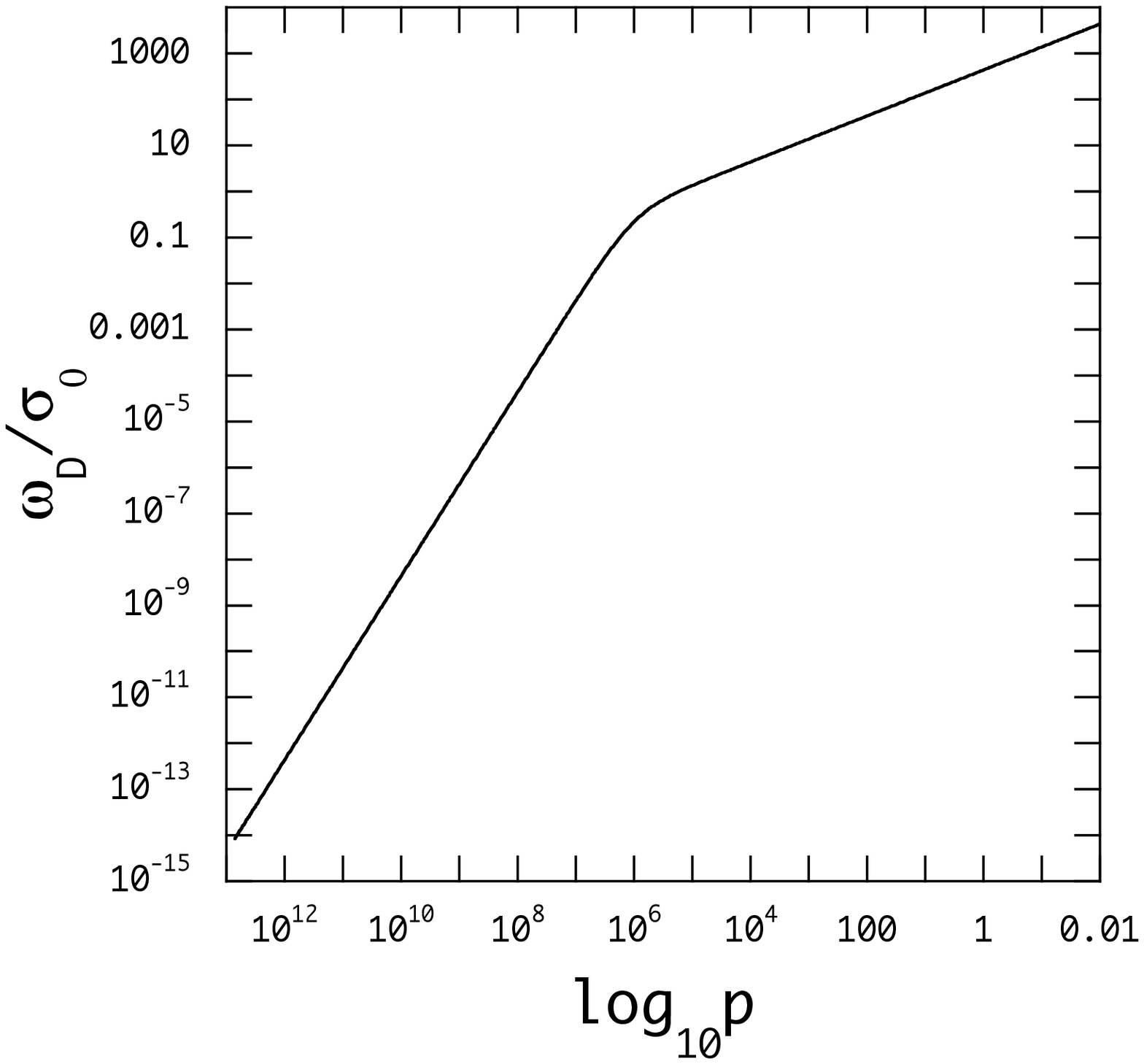}}
\hspace*{0.5cm}
\resizebox{0.45\columnwidth}{!}{
\includegraphics{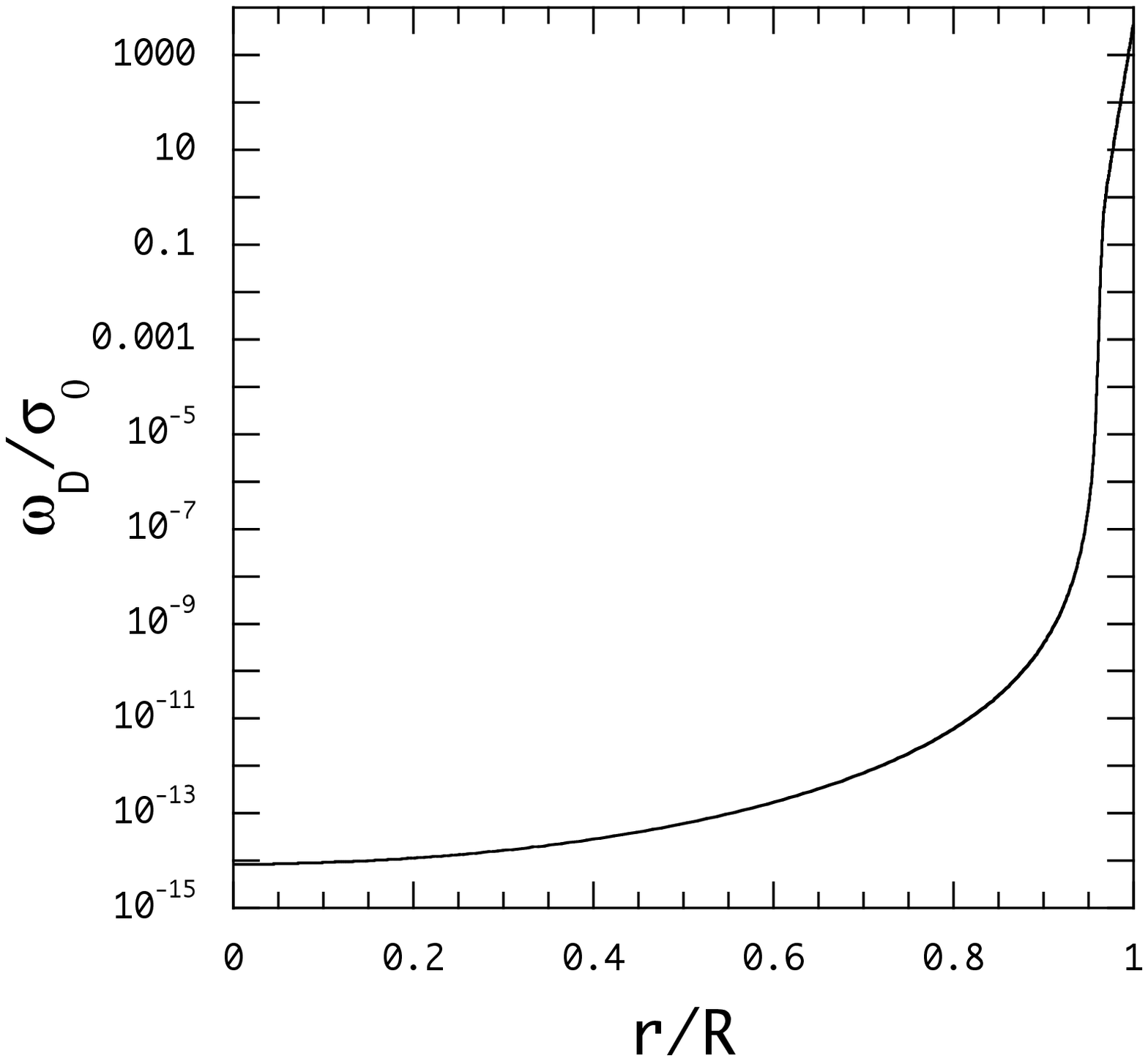}}
\caption{$\bar\omega_D$ as a function of $\log_{10}p$ (left panel) and of $r/R$ (right panel) for $\tau_*=1$day.
}
\label{fig:omegad}
\end{figure}

\begin{figure}
\resizebox{0.45\columnwidth}{!}{
\includegraphics{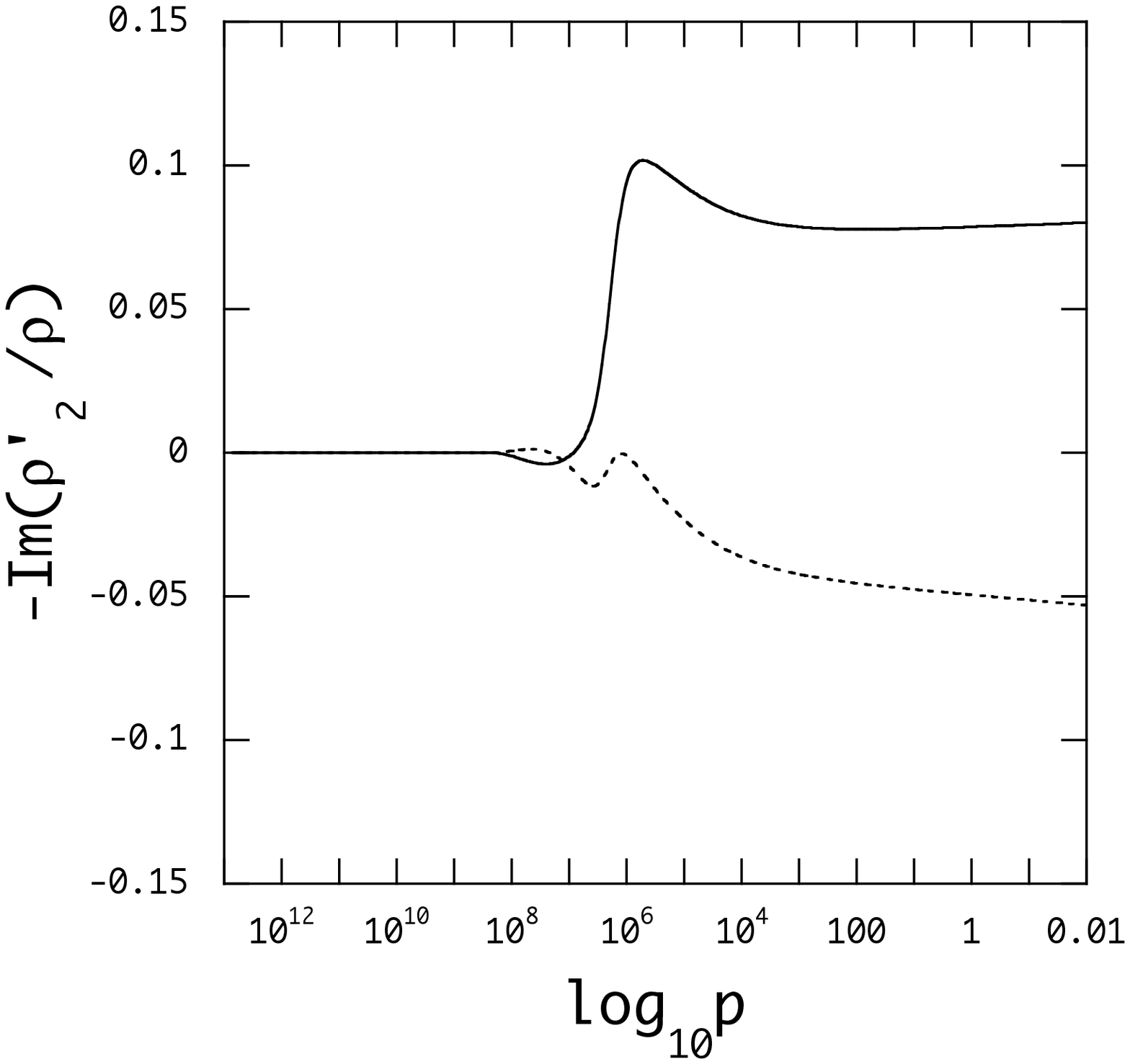}}
\hspace*{0.5cm}
\resizebox{0.45\columnwidth}{!}{
\includegraphics{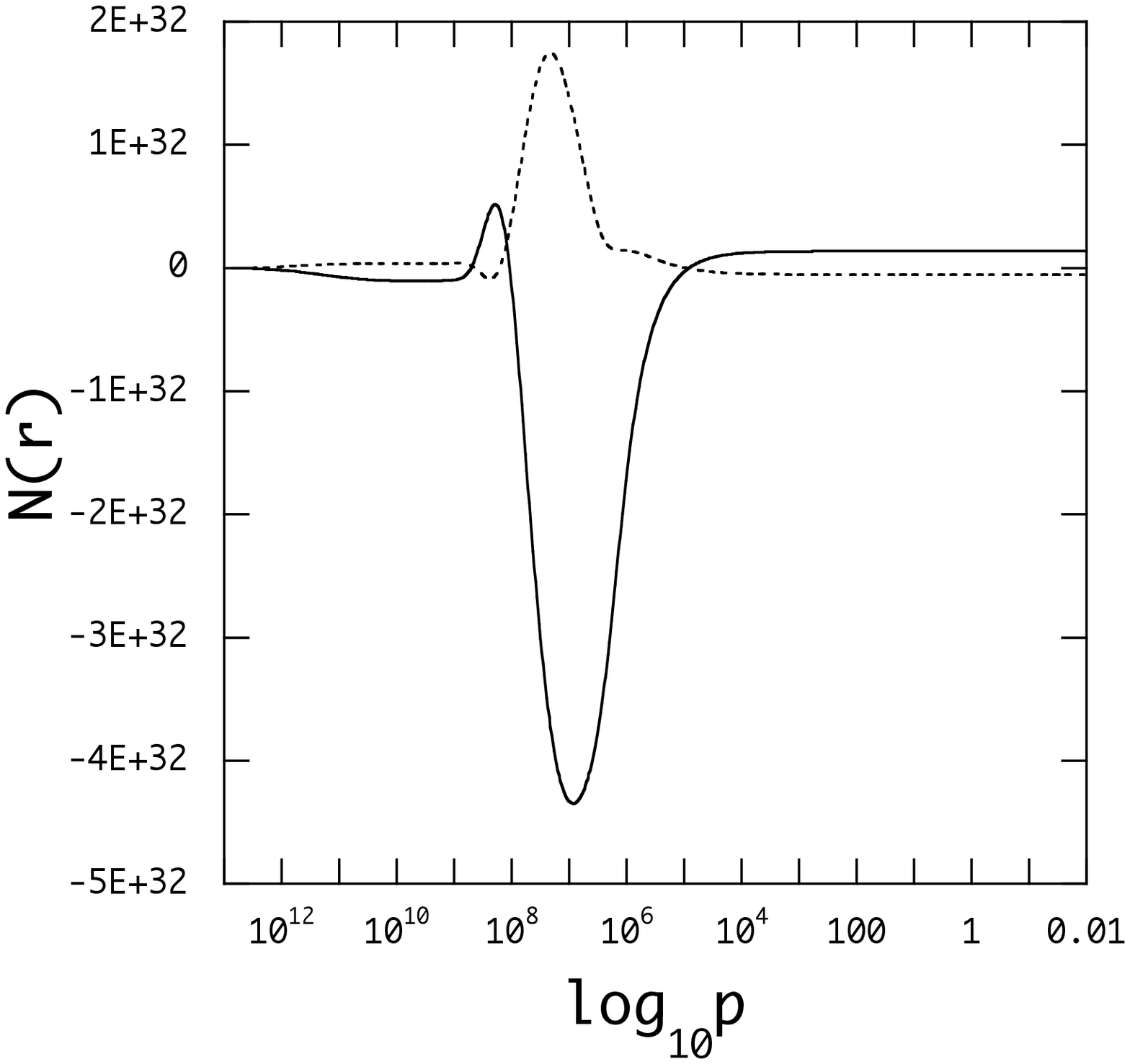}}
\caption{$-{\rm Im}(\rho^{\prime}_2/\rho)$ (left panel) and ${\cal N}(r)$ (right panel) as a function of $\log_{10}p$ for $\bar\Omega=0.1$ and $\tau_*=1$day,
where we have assumed $\pmb{j}_*=0$ and $\pmb{\psi}\not=0$.
The solid and dotted curves respectively correspond to the forcing frequency $\bar\omega=0.1$
and $0.0812297$.
}
\label{fig:rhoprime}
\end{figure}

Since we take no account of dissipative processes in the convective core except for radiative damping associated with Newtonian cooling, 
the results for the tidal torque obtained in this paper are not necessarily reliable, 
particularly for the case of $\pmb{\psi}\not=0$.
The perturbing tidal potential $\pmb{\psi}$ has substantial amplitudes in the core and hence the tidal responses to
$\pmb{\psi}$ in the core can be strongly affected by dissipative processes there and so is the tidal torque $\cal N$.
The Newtonian cooling in the envelope considered in this paper is controlled by the parameter $\omega_D$.
Fig. \ref{fig:omegad} plots $\bar\omega_D$ as a function of $\log_{10}p$ and $r/R$ for $\tau_*=1$day,
and shows that $\bar\omega_D$ increases by several orders of magnitudes within a geometrically thin layer near the bottom of the envelope
from $\bar\omega_D\sim 10^{-5}$ at $p\sim p_b$ to $\bar\omega_D\sim0.1$ at $p\sim p_*$.
For a given forcing frequency $\bar\omega$, strong tidal torque is produced in the layer of 
$\bar\omega\sim\bar\omega_D$, which occurs in this thin layer except in the limit of $\bar\omega\rightarrow 0$.
As equation (\ref{eq:tidaltorque}) indicates, the tidal response $\rho'_2$, particularly its imaginary part, plays an essential role to determine the tidal torque.
In Fig. \ref{fig:rhoprime}, the tidal response $-{\rm Im}(\rho_2^{\prime}/\rho)={\rm Im}(\rho_2^{\prime *}/\rho)$ 
and the cumulative tidal torque ${\cal N}(r)$ defined by
\be
{\cal N}(r)=\sqrt{3\pi\over 10}{GM_*\over a_*^3}\int_0^rdrr^4{\rm Im}[\rho'^*_2(r)]
\label{eq:tidaltorque_r}
\ee
are plotted for $\pmb{j}_*=0$ and $\pmb{\psi}\not=0$ for two forcing frequencies $\bar\omega=0.1$ and $0.0812297$, which respectively correspond to positive and negative $\cal N$, where we use
$\bar\Omega=0.1$ and $\tau_*=1$day.
As the figure indicates, significant changes of ${\rm Im}(\rho_2^{\prime *}/\rho)$
and $d{\cal N}/d\ln p$ occur in the region of $\bar\omega\sim\bar\omega_D$ and the tidal torque $\cal N$
is determined by the balance between positive and negative contributions of $d{\cal N}/d\ln p$ to ${\cal N}$ in the layer. 
The balance within this geometrically thin layer depends on the response $\rho'_2$ there and hence on the forcing frequency $\omega$.
Note that we find similar behavior of $-{\rm Im}(\rho_2^{\prime}/\rho)$ and ${\cal N}(r)$ also for the case of $\pmb{j}_*\not=0$ and $\pmb{\psi}=0$.
{\bf Because both forcing terms $\pmb{j}_*$ and $\pmb{\psi}$ in the perturbed entropy equation (\ref{eq:y6}) 
obtained under the Newtonian cooling approximation appear
with the same factor $1/(\rmi\omega+\omega_{\rm D})$, which is responsible for the deviation from adiabatic perturbations, 
the behaviors of the thermal responses to $\pmb{j}_*$ and $\pmb{\psi}$ become similar in the envelope.}
If we could correctly include the effects of dissipations in the convective core,
the results for the tidal torque $\cal N$ would be different from those computed in this paper,
particularly when we consider tidal responses to $\pmb{\psi}$ {\bf since the relation between the entropy perturbation and
the forcing $\pmb{\psi}$ in the core will be different from the relation we use for the envelope in this paper.}

\end{appendix}

\end{document}